\documentstyle[12pt]{article}
\newcommand {\smap}{\mbox{${\overline{\cal M}_{g,s}}({\mbox{M}\,,\,n})$}}
\newcommand {\NU}{\mbox{\Large $\nu$}}
\newcommand \gauss[1]{{\left[{#1}\right]}_{{\!}G}}

\newcommand {\dc}{\cal D}
\newcommand {\ps}{\sf p}
\newcommand {\gs}{\sf g}
\newcommand {\sfh}{\sf H}

\newcommand \corr[1]{\left\langle{#1}\right\rangle}

\newcommand {\lwedge}{\mbox{\Large $\wedge$}}

\newcommand {\Kappa}{\mbox{\Large $\kappa$}}

\newcommand {\beq}{\begin{eqnarray}}
\newcommand {\eeq}{\end{eqnarray}}
\newcommand {\beqs}{\begin{eqnarray*}}
\newcommand {\eeqs}{\end{eqnarray*}}
\newcommand {\del}{\partial}

\newcommand {\ca}{\cal A}
\newcommand {\bca}{\overline{\cal A}}

\newcommand \cm[2]{\left[ {#1}\, ,{#2} \right]}
\newcommand \ac[2]{\left\{ {#1}\, ,{#2}\right\}}
\newcommand \th[1]{{\Theta}_{#1}}

\newcommand {\bc}{\bf C}
\newcommand {\bz}{\bf Z}

\newcommand \ftu[1]{{\cal O}^{#1}}
\newcommand \ftbu[1]{\bar{{\cal O}}^{#1}}

\newcommand {\vpi}{\varpi}

\newcommand {\Om}{\Omega}

\newcommand \om[1]{{\omega}_{#1}}
\newcommand \omb[1]{\bar{{\omega}}_{#1}}

\newcommand {\kae}{\mbox{K{\"a}hler}}

\newcommand {\tr}{\mbox{Tr}}

\newcommand{\pr}{\hspace{\parindent}}

\begin{document}
\setlength{\oddsidemargin}{0cm}
\setlength{\baselineskip}{6.3mm}

\begin{titlepage}
    \begin{normalsize}
     \begin{flushright}
            { YITP/U-95-14} \\
            % {preliminary version} \\
            % April 1995
     \end{flushright}
    \end{normalsize}
    \begin{LARGE}
       \vspace{1cm}
       \begin{center}
        { Genus One Partition Function\\
            of the Calabi-Yau d-Fold\\
          embedded in ${{CP}^{d+1}}$\\
                  }
       \end{center}
    \end{LARGE}

   \vspace{5mm}

\begin{center}
           Katsuyuki S{\sc ugiyama}\footnote{E-mail address:
             ksugi@yisun1.yukawa.kyoto-u.ac.jp} \\
       \vspace{4mm}
                  {\it Uji Research Center, } \\
                  {\it Yukawa Institute for Theoretical Physics, } \\
                  {\it Kyoto University,
                  Uji 611, Japan} \\
       \vspace{1cm}

     \begin{large} ABSTRACT \end{large}
\par
  \end{center}
\begin{quote}
 \begin{normalsize}

For a one-parameter family of Calabi-Yau d-fold M embedded in
${{CP}^{d+1}}$, we consider a new quasi-topological field theory
${A^{\ast}}$(M)-model compared with the $A$(M)-model. The two point
correlators on the sigma model moduli space (the hermitian metrics)
are analyzed by the $A{A^{\ast}}$-fusion on the world sheet sphere. A
set of equations of these correlators turns out to be a non-affine
A-type Toda equation system for the d-fold M. This non-affine property
originates in the vanishing first Chern class of M. Using
the results of the $A{A^{\ast}}$-equation, we obtain a genus one
partition function of the sigma model associated to the M in the
recipe of the holomorphic anomaly. By taking an asymmetrical limit of
the complexified {\kae} parameters ${\bar{t}\rightarrow \infty}$ and $t$
is fixed, the ${A^{\ast}}$(M)-model part is decoupled and we can
obtain a partition function (or one point function of the operator
${{\cal O}^{(1)}}$ associated to a {\kae} form of M) of the
$A$(M)-matter coupled with the topological gravity at the stringy one
loop level. The coefficients of the series expansion with respect to
an indeterminate $q:={e^{2\pi i t}}$ are integrals of the top Chern
class of the vector bundle {\Large $\nu $} over the moduli space of
stable maps with definite degrees.

 \end{normalsize}
\end{quote}

\end{titlepage}
\vfill

\section{Introduction}

\pr
When one considers correlation functions of an N=2 non-linear sigma model
with a Calabi-Yau target space, they depend not only on the moduli space
of the Riemann surface, but also on the properties of the target Calabi-Yau
manifold (especially on the Calabi-Yau moduli spaces).
By twisting the N=2 non-linear sigma model {\cite{WIT,EY}}, one can obtain
two quasi-topological field theories (the A-model and the B-model)
{\cite{W}}, which describe two distinct Calabi-Yau
moduli spaces (a {\kae} structure moduli space and
a complex structure moduli space, respectively).

So far it was difficult to analyze non-perturbative corrections of the
A-model correlation functions. However by the discovery of the mirror
symmetry {\cite{GP,CLS}}
between the A(M)-model and the B(W)-model
for the mirror pairs (M, W), it is becoming possible to
investigate these corrections in the A-model {\cite{CDGP,Y}}.

Up to now, three point functions and d-point
functions on the genus 0 Riemann surface (i.e.
at the stringy tree level) for the Calabi-Yau d-folds
\cite{LSW,BV,NS,GMP,JN,sugi} and
partition functions in the higher genus ($g \geq 1$) (i.e.
at the stringy higher (more than or equal to one) loop level) for
Calabi-Yau 3-folds \cite{BCOV} have been investigated under mirror symmetries.
Now it may fairly be said that the analyses of the Calabi-Yau 3-folds under
the mirror symmetries have been established
{\cite{KT,F,KT2,CDFKM,CFKM,
BCDFHJQ,HKT}}.
%In this article, we study the
%A(M)-model correlators of some d-dimensional Calabi-Yau manifolds as
%mathematical physics applications under the mirror symmetry
%furthermore {\cite{NS,GMP,JN}}.

In this article, we take a genus one Riemann surface as a world sheet,
a (complex) d-dimensional Calabi-Yau target space M realized as a
hypersurface embedded in a projective space ${CP}^{d+1}$ and study
the properties of a partition function of the A(M)-model under the
mirror symmetry.

The rest of the paper is organized as follows. In section 2, we review
an $N=2$ supersymmetric non-linear sigma model with a Calabi-Yau
target space M and its topological version $A$(M)-model \cite{WIT,AM}.
Also another topological theory
${A^{\ast}}$(M)-model is introduced. In section 3,
a one-parameter family of the Calabi-Yau d-fold M and its mirror pair
W are explained. In order to study the hermitian metrics
${{\gs}_{l\bar{m}}}$ of the sigma model moduli space, we develop a
method $A{A^{\ast}}$-fusion there. This recipe is applied to our
one-parameter family of d-fold in section 4.
%The data of the
%$A$(M)-model three point functions on the sphere determine the metrics
%${\gs}$.
By using the data the author derived before, a set of equations for
the metrics ${{\gs}_{l\bar{m}}}$ is obtained. This equation system
turns out to be a non-affine A-type Toda equation system. We solve
this equation system explicitly and investigate the properties of the
metrics. Also the results obtained there are used to analyze a genus
one partition function for the d-fold. Section 5 provides a field
theoretical interpretation for the genus one partition function.
Section 6 is devoted to the conclusion and comments. In appendices,
several explanations and calculations omitted in the text are collected.

\section{Review of the A-Model}

\pr
In this section, we review an A(M)-model associated with a
d-dimensional Calabi-Yau manifold M. In addition, we introduce another
(quasi)-topological field theory ${A^{\ast}}$-model.

\subsection{$N=2$ Supersymmetric Non-linear Sigma Model}

\pr
To begin with, we introduce an $N=2$ supersymmetric non-linear sigma model
in two dimensions. Let $\Sigma$ be a Riemann surface and M be a
$d$-dimensional Calabi-Yau manifold. Locally one chooses
coordinate systems $({{X^i} ,{X^{\bar{\imath}}}})$ on M and
$({z, \bar{z}})$ on $\Sigma$ (where $X^i$ stands for a
set of the holomorphic coordinate system $({i=1,2,\cdots , d})$ ).
The $({{X^i} ,{X^{\bar{\imath}}}})$ are considered as mappings from
$\Sigma$ to M.
In this supersymmetric theory, there are fermionic pairs
$({{\psi^i_L},{\psi^{\bar{\imath}}_L}})$,\,
$({{\psi^i_R},{\psi^{\bar{\imath}}_R}})$\, where subscripts ``$L$'',``$R$''
stand for left-moving parts, right-moving parts on $\Sigma$ respectively.
Also superscripts ``$i$'', ``${\bar{\imath}}$'' mean that
${{\psi^i},{\psi^{\bar{\imath}}}}$ have values on
${{X^{\ast}}({T^{1,0} M})}$,
${{X^{\ast}}({T^{0,1} M})}$ respectively. With these fields, the Lagrangian
is written,
\beq
{L_0}{\!}&:=&{\!} \int_{\Sigma} {d^2}z \Biggl[ \frac{1}{2} g_{i \bar{\jmath}}
\left({ \del_z {X^i} \del_{\bar{z}} {X^{\bar{\jmath}}} +
\del_{\bar{z}} {X^i} \del_{z} {X^{\bar{\jmath}}}}\right) \nonumber \\
&+& \sqrt{-1} g_{i \bar{\jmath}} {\psi^{\bar{\jmath}}_L} {D_{\bar{z}}}
{\psi^{i}_L}+ \sqrt{-1} g_{i \bar{\jmath}} {\psi^{\bar{\jmath}}_R} {D_{z}}
{\psi^{i}_R}
+ R_{i \bar{\jmath} k \bar{l}}
{\psi^{i}_L}{\psi^{\bar{\jmath}}_L}{\psi^{k}_R}{\psi^{\bar{l}}_R}
\Biggr] \,\,\,.\label{eqn:orig}
\eeq
Here covariant derivatives
${D_{z}},{D_{\bar{z}}}$ are defined as,
\beqs
&&{D_z} {\psi^I_R} :=\frac{\del}{\del z} {\psi^I_R}+\frac{\del {X^J}}{\del z}
{\Gamma^{I}_{JK}} {\psi^K_R} \,\,\,,\\
&&{D_{\bar{z}}} {\psi^I_L} :=\frac{\del}{\del \bar{z}} {\psi^I_L}+
\frac{\del {X^J}}{\del \bar{z}}
{\Gamma^{I}_{JK}} {\psi^K_L} \,\,\,,
\eeqs
with ${\Gamma^{I}_{JK}}$ being the Levi-Civita connection of M and
the curvature ${R_{i \bar{\jmath} k \bar{l}}}$ is the Riemann tensor of M.
This Lagrangian system possesses an $N=2$ supersymmetry,
\beqs
\mbox{
\begin{tabular}{|c|c|c|c|c|} \hline
  & ${Q_R}$ & ${\widetilde{Q}_R}$ & ${Q_L}$ &
${\widetilde{Q}_L}$ \\ \hline
${X^i}$  & ${{\psi}_{R}^{i}}$ & $0$ & ${{\psi}_{L}^{i}}$ & $0$
\\ \hline
${X^{\bar{\imath}}}$  & $0$ & ${{\psi}_{R}^{\bar{\imath}}}$
& $0$ & ${{\psi}_{L}^{\bar{\imath}}}$
\\ \hline
${\psi^i_{L}}$  & ${-{\Gamma^{{i}}_{{j} {k}}}{\psi^{j}_R}
{\psi^{k}_{L}}}$ & $0$ & $0$ &
 ${\sqrt{-1} {\del_{z}}{X^{{i}}}}$  \\ \hline
${\psi^{\bar{\imath}}_L}$ & $0$ &
 ${- {\Gamma^{\bar{\imath}}_{\bar{\jmath} \bar{k}}}{\psi^{\bar{\jmath}}_R}
{\psi^{\bar{k}}_L}}$
 & ${\sqrt{-1} {\del_z}{X^{\bar{\imath}}}}$
& $0$ \\ \hline
${{\psi}_{R}^{i}}$  & $0$ & ${\sqrt{-1} {\del_{\bar{z}}{X^{i}}}}$ &
${-{\Gamma^{{i}}_{{j} {k}}}{\psi^{j}_R}
{\psi^{k}_{L}}}$ &
$0$ \\ \hline
${\psi^{\bar{\imath}}_R}$  &
${\sqrt{-1} {\del_{\bar{z}}{X^{\bar{\imath}}}}}$ &
$0$ & $0$ &
 ${- {\Gamma^{\bar{\imath}}_{\bar{\jmath} \bar{k}}}{\psi^{\bar{\jmath}}_R}
{\psi^{\bar{k}}_L}}$
\\ \hline
\end{tabular}
} \,\,\,,
\eeqs
\newline
where ${Q_L}$, ${\widetilde{Q}_L}$, ${Q_R}$, ${\widetilde{Q}_R}$
are super charges which generate super-transformations,
\beqs
&&{\delta_L}{\cal O}= \sqrt{-1}{\epsilon_L}\ac{Q_L}{\cal O}\,\,\,,\\
&&{\tilde{\delta}_L}{\cal O}= \sqrt{-1}{\tilde{\epsilon}_L}
\ac{\widetilde{Q}_L}{\cal O}\,\,\,,\\
&&{\delta_R}{\cal O}= \sqrt{-1}{\epsilon_R}\ac{Q_R}{\cal O}\,\,\,,\\
&&{\tilde{\delta}_R}{\cal O}= \sqrt{-1}{\tilde{\epsilon}_R}
\ac{\widetilde{Q}_R}{\cal O}\,\,\,,\\
&&\,\,\,\,
{\epsilon_L}\,,\,{\tilde{\epsilon}_L}\,,\,{\epsilon_R}\,,\,
{\tilde{\epsilon}_R}\,;\,\mbox{fermionic parameters}\,\,\,.
\eeqs
These charges satisfy the following anti-commutation relations,
\beqs
&&{Q^{2}_{L}}={\tilde{Q}^{2}_{L}}=
{Q^{2}_{R}}={\tilde{Q}^{2}_{R}}=0\,\,\,,\\
&&\ac{Q_L}{Q_R}=\ac{\tilde{Q}_L}{\tilde{Q}_R}=0\,\,\,,\\
&&\ac{\tilde{Q}_L}{Q_R}=\ac{{Q}_L}{\tilde{Q}_R}=0\,\,\,,\\
&&\ac{Q_L}{\tilde{Q}_L}={{\del}_z}\,\,,\,\,
\ac{{Q}_R}{\tilde{Q}_R}={\bar{\del}_{\bar{z}}}
\,\,\,.
\eeqs
%%%%##################################
\subsection{Topological Sigma Model}

\pr
In order to obtain topological versions of the $N=2$ non-linear
sigma model, let us consider the alternation of bundles on which fermions
take values. We change spins of fermions by an amount depending on their
$U(1)$ charges.
As a result, fermions take values not on spin bundles but on
(anti)-canonical bundles. Especially we consider two cases;\\
\mbox{}\\
$\bullet$ Case $I$ ({$A$(M)-Model})\\
In this case, we change the bundles on which fermions take values
and rename the resulting fields as following,
\beqs
%\]
&&\left\{
\begin{array}{ccccccr}
{\psi^{i}_L} & : & {K^{1/2}}\otimes {X^{\ast}}({T^{1,0}}M) & \rightarrow
 & \chi^{i} & : & {K^0} \otimes {X^{\ast}}({T^{1,0}}M) \\
{\psi^{\bar{\imath}}_L} & : & {K^{1/2}}\otimes {X^{\ast}}({T^{0,1}}M)
& \rightarrow & \rho^{\bar{\imath}}_z & : & {K^1}
\otimes {X^{\ast}}({T^{0,1}}M) \\
{\psi^{i}_R} & : & {\bar{K}^{1/2}}\otimes {X^{\ast}}({T^{1,0}}M) & \rightarrow
 & \rho^{i}_{\bar{z}} & : & {\bar{K}^1} \otimes {X^{\ast}}({T^{1,0}}M) \\
{\psi^{\bar{\imath}}_R} & : & {\bar{K}^{1/2}}\otimes {X^{\ast}}({T^{0,1}}M)
& \rightarrow  & \chi^{\bar{\imath}} & : & {\bar{K}^0}
\otimes{X^{\ast}}({T^{0,1}}M)
\end{array}
\right.\,\,\,,
\eeqs
where the $K$ is a canonical bundle.
\newline
$\bullet$ Case $II$ ({${A^{\ast}}$(M)-Model})\\
Similarly to the case (I), we can perform an opposite twisting,
\beqs
%\]
&&\left\{
\begin{array}{ccccccr}
{\psi^{i}_L} & : & {K^{1/2}}\otimes {X^{\ast}}({T^{1,0}}M) & \rightarrow
 & \bar{\rho}^{i}_z & : & {K^1}\otimes {X^{\ast}}({T^{1,0}}M) \\
{\psi^{\bar{\imath}}_L} & : & {K^{1/2}}\otimes {X^{\ast}}({T^{0,1}}M)
& \rightarrow & \bar{\chi}^{\bar{\imath}} & : &
{K^0} \otimes {X^{\ast}}({T^{0,1}}M) \\
{\psi^{i}_R} & : & {\bar{K}^{1/2}}\otimes {X^{\ast}}({T^{1,0}}M) & \rightarrow
 & \bar{\chi}^{i}  & : & {\bar{K}^0} \otimes {X^{\ast}}({T^{1,0}}M) \\
{\psi^{\bar{\imath}}_R} & : & {\bar{K}^{1/2}}\otimes {X^{\ast}}({T^{0,1}}M)
& \rightarrow  & \bar{\rho}^{\bar{\imath}}_{\bar{z}} & : &
{\bar{K}^1}\otimes {X^{\ast}}({T^{0,1}}M)
\end{array}
\right.\,\,\,.
%\]
%\[
\eeqs
We call the former model as ${A}$(M)-model and the latter as
${A^{\ast}}$(M)-model.
As topological theories, the super charges ${\widetilde{Q}_L}$ and
${Q_R}$ are combined into a BRST charge ${Q^{(+)}}$,
\[
{{Q^{(+)}}:={{Q}_L}
+{\widetilde{Q}_R}}\,\,\,,
\]
in the $A$(M)-model. On the other hand,
a BRST charges ${Q^{(-)}}$ is constructed from
the remaining ones ${Q_L}$ and ${\widetilde{Q}_R}$ in the
${A^{\ast}}$(M)-model,
\[
{{Q^{(-)}}:={\widetilde{Q}_L}+{{Q}_R}}\,\,\,.
\]
Local observables of the $A$(M)-model, the
${A^{\ast}}$(M)-model are defined as elements of the BRST
cohomologies with respect to the BRST charges
${Q^{(+)}}$, ${Q^{(-)}}$ respectively.

Firstly in the $A$(M)-model, local observables
%${\ftu{(i)}}$
%\,({\ftbu{({\bar{\jmath}})}})$
are functionals of the fields
%\,({A_{t^{\ast}}})$
$({X^{i}},{X^{\bar{\jmath}}},
{\chi^{i}},{\chi^{\bar{\jmath}}})$.
%\,({\,{$({X^{i}},{X^{\bar{\jmath}}},
%{\bar{\chi}^{i}},{\bar{\chi}^{\bar{\jmath}}})$}\,}).
Considering the correspondence between the de~Rham cohomology and the
A-model BRST cohomology, we associate an arbitrary de~Rham cohomology
element $\omega $ to a physical observables ${{\phi_A}[{\omega}]}$
in the $A$(M)-model,
\beqs
&&\omega ={\omega_{{i_1}\cdots{i_p}{\bar{\jmath_1}}\cdots{\bar{\jmath_q}}}}
d{X^{i_1}}{\wedge}\cdots{\wedge}d{X^{i_p}}{\wedge}
d{X^{\bar{\jmath}_1}}{\wedge}\cdots{\wedge}d{X^{\bar{\jmath}_q}}\in
{H^{p,q}_{d}}(M)\,\,\,, \\
&&\hspace{2cm}(\,d{\omega}=0\,\,,\,\,{\omega}\sim\omega +d\nu\,)\,\,\,,\\
\leftrightarrow  &&{\phi}_A [{\omega}]
={\omega_{{i_1}\cdots{i_p}{\bar{\jmath_1}}\cdots{\bar{\jmath_q}}}}
{\chi^{i_1}}\cdots{\chi^{i_p}}{\chi^{\bar{\jmath}_1}}
\cdots {\chi^{\bar{\jmath}_q}}
\,\,\,,\\
&&\hspace{2cm}(\,\delta {\phi_A} [{\omega}])=0 \,)\,\,\,.
\eeqs
Especially a relation is satisfied,
\[
\{{{Q^{(+)}},{{\phi_A} [{\omega}]}}\}= {\phi_A} [{d{\omega}}] \,\,\,.
\]

Secondly ${A^{\ast}}$(M)-model observables are constructed from
the fields $({X^{i}},{X^{\bar{\jmath}}},
{\overline{\chi}^{i}},{\overline{\chi}^{\bar{\jmath}}})$.
In similar to the $A$(M)-model case, we can define a physical
operator ${{{\phi}_{A^{\ast}}}[{\omega}]}$ in the
${A^{\ast}}$(M)-model for each cohomology element ${\omega}$,
\beqs
&&\omega ={\omega_{{i_1}\cdots{i_p}{\bar{\jmath_1}}\cdots{\bar{\jmath_q}}}}
d{X^{i_1}}{\wedge}\cdots{\wedge}d{X^{i_p}}{\wedge}
d{X^{\bar{\jmath}_1}}{\wedge}\cdots{\wedge}d{X^{\bar{\jmath}_q}}\in
{H^{p,q}_{d}}(M)\,\,\,, \\
&&\hspace{2cm}(\,d{\omega}=0\,\,,\,\,{\omega}\sim\omega +d\nu\,)\,\,\,,\\
\leftrightarrow  &&{\phi}_{A^{\ast}} [{\omega}]
={\omega_{{i_1}\cdots{i_p}{\bar{\jmath_1}}\cdots{\bar{\jmath_q}}}}
{\overline{\chi}^{i_1}}\cdots{\overline{\chi}^{i_p}}
{\overline{\chi}^{\bar{\jmath}_1}}
\cdots {\overline{\chi}^{\bar{\jmath}_q}}
\,\,\,,\\
&&\hspace{2cm}(\,\delta {\phi_{A^{\ast}}} [{\omega}])=0 \,)\,\,\,.
\eeqs
Then a relation is satisfied,
\[
\{{{Q^{(-)}},{{\phi_{A^{\ast}}} [{\omega}]}}\}
= {\phi_{A^{\ast}}} [{d{\omega}}] \,\,\,.
\]

\section{The ${A{A^{\ast}}}$-Fusion and the Two Point Functions}

\pr
In this section, we investigate the mixing between the holomorphic
part and the anti-holomorphic one  (the hermitian metrics)
in the Calabi-Yau non-linear sigma model.

\subsection{The One-Parameter Model in ${{CP}^{d+1}}$}

\pr
Throughout this article, we take a one-parameter family of Calabi-Yau
d-fold M realized as a zero locus of a hypersurface embedded in a
projective space ${{CP}^{d+1}}$,
\beq
&&M\,;\,{\ps}={X_1^{d+2}}+{X_2^{d+2}}+\cdots +{X_{d+2}^{d+2}}\nonumber
\\
&&\hspace{2cm}-(d+2)\psi ({X_1}{X_2}\cdots {X_{d+2}})=0
\,\,\,in \,\,\, {CP}^{d+1}\,\,\,,\label{eqn:CY}
\eeq
as a target space in the N=2 non-linear sigma model. Hodge numbers ${h^{p,q}}$
of
this d-fold are calculated as \cite{Dais},
\beq
&&{h^{p,q}}={\delta_{p,q}}\,\,\,,\,\,\,
(0 \leq p \leq d\,,\, 0 \leq q \leq d \,,\, p+q \neq d)\,\,\,,\nonumber \\
&&{h^{d-p,p}}={\delta_{2p,d}}+
{\sum^{p}_{i=0}}{{(-1)}^{i}}
\left(
\begin{array}{c}
 d+2 \\
 i
\end{array}
\right)
\cdot
\left(
\begin{array}{c}
 (p+1-i)(d+1)+p \\
  d+1
\end{array}
\right)\,\,\,,\,\,\,
(0 \leq p \leq d )\,\,\,.\label{eqn:hodge}
\eeq
Especially a Euler number ${\chi}(M)$ can be written down,
\[
{\chi}(M)=\frac{1}{N}\left\{{
{{({1-N})}^N}-1+{N^2}
}\right\} \,\,\,,\,\,\,N:=d+2\,\,\,.
\]
A mirror manifold W paired with this M is constructed as a orbifold
divided by some maximally invariant discrete group
$G={{\left({{\bz}_{d+2}}\right)}^{
(d+1)}}$,
\beqs
\mbox{W}\,;\,\{{\ps =0}\}/G \,\,\,.
\eeqs
This discrete group acts on the coordinate
$({{X_1}\,,\,{X_2}\,,\,\cdots \,,\,{X_{d+1}}\,,\,{X_{d+2}}})$ as \\
$({{{\tilde{\alpha}}^{a_1}}{X_1}\,,\,
{{\tilde{\alpha}}^{a_2}}{X_2}\,,\,\cdots \,,\,
{{\tilde{\alpha}}^{a_{d+1}}}{X_{d+1}}\,,\,
{{\tilde{\alpha}}^{a_{d+2}}}{X_{d+2}}})$  with ${\tilde{\alpha} :={\exp {\!}
\left({\frac{2\pi i}{d+2}}\right)}}$,
\beqs
({{a_1}\,,\,{a_2}\,,\,{a_3}\,,\,
\cdots \,,\,{a_d}\,,\,{a_{d+1}}\,,\,{a_{d+2}}})=
\left\{
\begin{array}{c}
({1\,,\,0\,,\,0\,,\,\cdots \,,\,0\,,\,0\,,{d+1}})\,\,\,,\\
({0\,,\,1\,,\,0\,,\,\cdots \,,\,0\,,\,0\,,{d+1}})\,\,\,,\\
({0\,,\,0\,,\,1\,,\,\cdots \,,\,0\,,\,0\,,{d+1}})\,\,\,,\\
\cdots \\
({0\,,\,0\,,\,0\,,\,\cdots \,,\,1\,,\,0\,,{d+1}})\,\,\,,\\
({0\,,\,0\,,\,0\,,\,\cdots \,,\,0\,,\,1\,,{d+1}})\,\,\,.
%({0\,,\,0\,,\,0\,,\,\cdots \,,\,1\,,\,0\,,{d+1}})\,\,\,,
\end{array}
\right.\,\,\,
\eeqs
When one thinks about the Hodge structure of the $G$-invariant parts
of the cohomology group ${H^{d}}$(W),
their Hodge numbers are written
as,
\[
{h^{d,0}}={h^{d-1,1}}=\cdots =
{h^{1,d-1}}={h^{d,0}}=1\,\,\,.
\]

\subsection{The Hermitian Metrics}

\pr
Let us take a set of elements $\{{\omega_l}\} \in
{H^{l,l}}({\mbox{M}})$
which can be
obtained from a {\kae} form
${J:={\omega_1}\in {H^{1,1}}({\mbox{M}})}$.
Each dimension of the primary vertical subspace
${{H^{l,l}}({\mbox{M}})}$ ${(0 \leq l \leq d)}$ is given as,
\beqs
\dim {H^{l,l}}({\mbox{M}})=
\left\{
\begin{array}{lcl}
1+{\displaystyle \sum^{\frac{d}{2}}_{i=0}}
{{(-1)}^{i}}
  \left(
    \begin{array}{c}
     d+2 \\
      i
    \end{array}
  \right)
\cdot
  \left(
    \begin{array}{c}
     \left({\frac{d}{2} +1-i}\right)\cdot (d+1)+\frac{d}{2} \\
      d+1
    \end{array}
  \right)
  & , & ({l=\frac{d}{2}\,\, \mbox{and $d$ is even}}) \,\,\,, \\
1 & , & ({\mbox{otherwise}}) \,\,\,.
\end{array}
\right.
\eeqs
But the dimension of the subspaces ${{H^{l,l}_J}({\mbox{M}})}$
obtained from a {\kae} form
${J}$
is written down,
\[
\dim {H^{l,l}_J}({\mbox{M}})=1\,\,\,,\,\,\,({0\leq l \leq d}) \,\,\,.
\]
{}From these cohomology elements ${\omega_l}\in {H^{l,l}_J}$(M), we can
construct physical observables
${{{\cal O}^{(l)}}:={\phi_A}[{\omega_l}]}$ in the $A$(M)-model and
${{\bar{\cal O}^{(\bar{l})}}:=
{\phi_{A^{\ast}}}[{\omega_l}]}$ in the ${A^{\ast}}$(M)-model.
These elements describe properties of the holomorphic part and the
anti-holomorphic one in the target space M. Let us take each one element
from the $A$-model part and the ${A^{\ast}}$-model one.
These two elements ${{\cal O}^{(l)}}$, ${\bar{\cal
O}^{(\bar{m})}}$ are combined into one correlator
${{{\gs}_{l\bar{m}}}:=
\corr{{\bar{\cal O}^{(\bar{m})}}{{\cal O}^{(l)}}}}$.
These correlators are different from the topological metrics in the
$A$(M)-model. These ${{\gs}_{l\bar{m}}}$ are hermitian as matrices
and are called the hermitian metrics.

Now we explain the definition of the expectation value in the above
formula. Firstly note that the $A$(M)-model Lagrangian ${L_A}$ can be
written as,
\beq
{L_A}={\int_{\Sigma}}{d^2}z \ac{Q^{(+)}}{V^{(-)}}-
\sqrt{-1}{\int_{\Sigma}}{X^{\ast}}(e)\,\,\,, \label{eqn:modela}
\eeq
where ${Q^{(+)}}$ is a BRST charge of the $A$(M)-model and ${V^{(-)}}$
is defined as,
\beqs
&&{V^{(-)}}:=\sqrt{-1}{g_{i\bar{\jmath}}}\,
({{\rho^{\bar{\jmath}}_{z}}{\del_{\bar{z}}}{X^{i}}+
{\rho^{i}_{\bar{z}}}{\del_{{z}}}{X^{\bar{\jmath}}}
})\,\,\,,\\
&&\,\,\,\,\,{g_{i\bar{\jmath}}}\,:\,
\mbox{a metric of the Calabi-Yau d-fold M}\,\,\,.
\eeqs
Also the $e$ is a {\kae} form of M,
\[
e:=\sqrt{-1}{g_{i\bar{\jmath}}}\,
d{X^i}\wedge d{X^{\bar{\jmath}}}\,\,\,,
\]
and the second term of ${L_A}$,
\beqs
{\int_{\Sigma}}{X^{\ast}}(e)=
{\int_{\Sigma}}\sqrt{-1}{g_{i\bar{\jmath}}}\,
({{\del_z}{X^i}{\del_{\bar{z}}}{X^{\bar{\jmath}}}
-{\del_{\bar{z}}}{X^i}{\del_z}{X^{\bar{\jmath}}}}){d^2}z \,\,\,,
\eeqs
is the integral of the pullback of the {\kae} form $e$.

Secondly the Lagrangian ${L_{A^{\ast}}}$ of the ${A^{\ast}}$(M)-model
can be expressed as,
\beq
{L_{A^{\ast}}}={\int_{\Sigma}}{d^2}z \ac{Q^{(-)}}{V^{(+)}}+
\sqrt{-1}{\int_{\Sigma}}{X^{\ast}}(e)\,\,\,,\label{eqn:modelba}
\eeq
where ${Q^{(-)}}$ is a BRST charge of the ${A^{\ast}}$(M)-model
and ${V^{(+)}}$
is defined as,
\beqs
{V^{(+)}}:=\sqrt{-1}{g_{i\bar{\jmath}}}\,
({{\bar{\rho}^{\bar{\jmath}}_{\bar{z}}}{\del_{{z}}}{X^{i}}+
{\bar{\rho}^{i}_{{z}}}{\del_{\bar{z}}}{X^{\bar{\jmath}}}
})\,\,\,.
\eeqs
The second term in the ({\ref{eqn:modela}})({\ref{eqn:modelba}})
has a peculiar property,
\beq
{\int_{\Sigma}}{X^{\ast}}(e)=
{\int_{\Sigma}}
\ac{{Q}_R}{\cm{\widetilde{Q}_L}{\sqrt{-1}{g_{i\bar{\jmath}}}\,{\psi^{i}_{L}}
{\psi^{\bar{\jmath}}_{R}} }}
+{\int_{\Sigma}}
\ac{\widetilde{Q}_R}{\cm{{Q}_L}{\sqrt{-1}{g_{i\bar{\jmath}}}\,{\psi^{i}_{R}}
{\psi^{\bar{\jmath}}_{L}} }} \,\,\,,\label{eqn:kae}
\eeq
where ${{Q_L}\,,\,{\widetilde{Q}_L}\,,\,{Q_R}\,,\,{\widetilde{Q}_R}}$
are super charges and
${({{\psi^{i}_{L}}\,,\,{\psi^{\bar{\jmath}}_{L}}})}$,
${({{\psi^{i}_{R}}\,,\,{\psi^{\bar{\jmath}}_{R}}})}$ are fermion pairs
of the original N=2 non-linear sigma model.
The operators ${{g_{i\bar{\jmath}}}{\psi^{i}_{L}}
{\psi^{\bar{\jmath}}_{R}}}$ and ${{g_{i\bar{\jmath}}}{\psi^{i}_{R}}
{\psi^{\bar{\jmath}}_{L}}}$ in the integrals are translated in the
$A$(M)-model and/or ${A^{\ast}}$(M)-model after twisting,
\beqs
\mbox{
\begin{tabular}{|c|c|c|} \hline
\mbox{Operators} & $A$-model & ${A^{\ast}}$-model \\ \hline
${{g_{i\bar{\jmath}}}{\psi^{i}_{L}}
{\psi^{\bar{\jmath}}_{R}}}$ &
${{g_{i\bar{\jmath}}}{\chi^{i}}
{\chi^{\bar{\jmath}}}}$ &
${{g_{i\bar{\jmath}}}{\bar{\rho}^{i}_{z}}
{\bar{\rho}^{\bar{\jmath}}_{\bar{z}}}}$ \\ \hline
${{g_{i\bar{\jmath}}}{\psi^{i}_{R}}
{\psi^{\bar{\jmath}}_{L}}}$ &
${{g_{i\bar{\jmath}}}{{\rho}^{i}_{\bar{z}}}
{{\rho}^{\bar{\jmath}}_{{z}}}}$ &
${{g_{i\bar{\jmath}}}{\bar{\chi}^{i}}
{\bar{\chi}^{\bar{\jmath}}}}$ \\ \hline
\end{tabular}
} \,\,\,.
\eeqs
\newline
For the $A$-model, the operator ${{\ftu{(1)}}:=
{g_{i\bar{\jmath}}}{\chi^{i}}{\chi^{\bar{\jmath}}}}$ is a BRST
observable associated to the {\kae} form $e$, but the other operator
${{g_{i\bar{\jmath}}}{\rho^{i}_{\bar{z}}}{\rho^{\bar{\jmath}}_{z}}}$
is not a BRST observable and not a physical one. However the second
term
${\ac{\widetilde{Q}_R}{\cm{{Q}_L}{\sqrt{-1}{g_{i\bar{\jmath}}}
\,{\rho^{i}_{\bar{z}}}
{\rho^{\bar{\jmath}}_{z}} }}}$ in ({\ref{eqn:kae}})
works trivially in physical situation
except for the cases when one must consider some contributions from
the boundary of the moduli space because ${{Q^{(+)}}={{Q}_L}+
{\widetilde{Q}_R}}$ is a BRST charge of $A$-model.

Conversely the operator ${{\ftbu{(\bar{1})}}:=
{g_{i\bar{\jmath}}}{\bar{\chi}^{i}}{\bar{\chi}^{\bar{\jmath}}}}$
turns out to be a physical one associated with the {\kae} form $e$ and
the other operator ${{g_{i\bar{\jmath}}}
{\bar{\rho}^{i}_{{z}}}{\bar{\rho}^{\bar{\jmath}}_{\bar{z}}}}$ becomes
unphysical for the ${A^{\ast}}$-model case. In similar to the
$A$-model case, the first term
${\ac{{Q}_R}{\cm{\widetilde{Q}_L}{\sqrt{-1}{g_{i\bar{\jmath}}}
\,{\bar{\rho}^{i}_{z}}
{\bar{\rho}^{\bar{\jmath}}_{\bar{z}}} }}}$ in ({\ref{eqn:kae}}) works trivially
in usual
physical situations because ${{Q^{(-)}}={\widetilde{Q}_L}+{{Q}_R}}$ is a
BRST operator of ${A^{\ast}}$-model.

Next we think about the meaning of the twisting. This twisting method
can also be understood in the Lagrangian formalism. To begin with, we
consider a Lagrangian ${L_0}$ ({\ref{eqn:orig}})
of the original N=2 non-linear sigma
model. The (quasi)-topological field theories (the $A$-model and the
${A^{\ast}}$-model) are obtained by introducing a fermion number
current ${\sf J}$ (one-form on the Riemann surface $\Sigma $),
\beqs
&&{L_A}={L_0}+{\int_{\Sigma}}{\sf J}\wedge \left({\frac{i}{2}\varsigma }\right)
\,\,\,,\,\,\,({A-{\mbox{model}}})\,\,\,,\\
&&{L_{A^{\ast}}}
={L_0}+{\int_{\Sigma}}{\sf J}\wedge \left({\frac{-i}{2}\varsigma }\right)
\,\,\,,\,\,\,({{A^{\ast}}-{\mbox{model}}})\,\,\,,
\eeqs
which is coupled to the spin connection $\varsigma $ on the Riemann
surface ${\Sigma}$. By this recipe, the spins of fermions are changed
by an amount depending on their U(1) charges and then fermions come to
take values not on spin bundles but on (anti)-canonical bundles. The
objects we want to obtain are the hermitian metrics ${{\gs}_{l\bar{m}}}
:=\corr{\ftbu{({\bar{m}})}\ftu{(l)}}$.
Because operators ${\ftu{(l)}}$, ${\ftbu{({\bar{m}})}}$ are observables of
the $A$-model, ${A^{\ast}}$-model respectively, it is necessary to
merge these two models into one theory. In order to fulfill our
purpose, we take a genus 0 Riemann surface as a world sheet ${\Sigma}$
constructed from two hemi-spheres ${\Sigma_L}$ and ${\Sigma_R}$,
\beqs
&&{\Sigma}={\Sigma_L} \cup {\Sigma_R}\,\,\,,\\
&&{\Sigma_L}\cap {\Sigma_R} \cong {S^1}\,\,\,,
\eeqs
where subscripts $L$ and $R$ stand for the left, right part of
${\Sigma}$ respectively. We put a topological theory on ${\Sigma_L}$,
an anti-topological one on ${\Sigma_R}$ and connect them smoothly on
${\Sigma}$. A Lagrangian of the resulting theory is defined as,
\beq
&&L({t\,,\,\bar{t}})=
{L_0}+{\int_{\Sigma}}{\sf J}\wedge {\sf A} \nonumber \\
&&\hspace{2cm} +t{\int}
\ac{{Q}_R}{\cm{\widetilde{Q}_L}{\sqrt{-1}{g_{i\bar{\jmath}}}\,{\psi^{i}_{L}}
{\psi^{\bar{\jmath}}_{R}} }} \nonumber \\
&&\hspace{2cm} + \bar{t}{\int}
\ac{\widetilde{Q}_R}{\cm{{Q}_L}{\sqrt{-1}{g_{i\bar{\jmath}}}\,{\psi^{i}_{R}}
{\psi^{\bar{\jmath}}_{L}} }} \,\,\,,\label{eqn:connect}
\eeq
where ${\sf A}$ is a U(1) gauge connection on ${\Sigma}$. This
${\sf A}$ becomes ${\frac{i}{2}\varsigma }$ in the far-right and turns
into ${\frac{-i}{2}\varsigma }$ in the far-left on ${\Sigma}$ smoothly.
The third and fourth terms on the right hand side in
({\ref{eqn:connect}})
 are perturbation terms in order to look into the response for the marginal
perburbations associated with the {\kae} form $e$. Using this
Lagrangian, we can write a definition of the hermitian metrics,
\beqs
{{\gs}_{l\bar{m}}}&=&\corr{\ftbu{({\bar{m}})}{\ftu{(l)}}}\\
&:=& \int {\!}{\cal D}[{X, \psi}]\, \ftbu{({\bar{m}})}\ftu{(l)}
{{e}^{-L({t\,,\,\bar{t}})}}\,\,\,.
\eeqs
By analyzing them in the operator formalism, we obtain an equation
({${A{A^{\ast}}}$-equation}) for these correlators
${{\gs}_{l\bar{m}}}$,
\beq
&&{\del_{\bar{t}}}\left({{\gs}{\del_t}{{\gs}^{-1}}}\right)=
\cm{C_t}{{\gs}{\bar{C}_{\bar{t}}}{{\gs}^{-1}}}\,\,\,,\label{eqn:fusion} \\
&& {{\left({C_t}\right)}_{lm}}=
\corr{\ftu{(m)}{\Big|}{\ftu{(1)}}{\Big|}{\ftu{(l)}}}\,\,\,,\nonumber \\
&& {{\left({\bar{C}_{\bar{t}}}\right)}_{\bar{l}\bar{m}}}=
\corr{\ftbu{(\bar{m})}{\Big|}{\ftbu{(\bar{1})}}{\Big|}
{\ftbu{(\bar{l})}}}\,\,\,,\nonumber
\eeq
where ${\left({C_t}\right)}$, ${\left({\bar{C}_{\bar{t}}}\right)}$ are
three point functions containing operators ${\ftu{(1)}}$,
${\ftbu{(\bar{1})}}$ for the $A$(M)-model, ${A^{\ast}}$(M)-model
respectively. (The derivation of this equation is explained in the
Appendix A).

The hermitian metrics are determined by three point correlators of the
$A$ $({A^{\ast}})$-models. The author obtained these couplings of the
${A}$(M)-model for the Calabi-Yau d-fold \cite{sugi}.
(A short review of the derivation of these correlators are explained
in the Appendix B). In the next section, we apply this formula to the
Calabi-Yau d-fold case ({\ref{eqn:CY}})
in order to obtain expressions of the metrics.

\section{Application of the ${A{A^{\ast}}}$-Equation}

\pr
In this section, we apply the $A{A^{\ast}}$-equation to the Calabi-Yau
d-fold M (\ref{eqn:CY}) and analyze properties of the hermitian
metrics ${{\gs}_{l\bar{m}}}$. Also by using these results, a genus one
partition function can be obtained.

\subsection{The Hermitian Metrics of the Calabi-Yau d-Fold}

\pr
When we take the set of the observables ${\{{\ftu{(l)}}\}}$,
${\left({\ftu{(l)} \in
{H^{l,l}_J}({\mbox{M}})\,\,;\,\,
({l=0,1,\cdots ,d})
}\right)}$
for the d-fold M as a basis, the
structure constant of the operator product between ${\ftu{(1)}}$ and
${\ftu{(l)}}$ can be represented as a matrix ${C_t}$ \cite{sugi},
\beq
&&{C_t}:= \left(
\begin{array}{ccccccc}
 0 & {\Kappa_0} &  &  &  &  & \mbox{\Large $O$} \\
   & 0 & {\Kappa_1}  &  &  &  &  \\
   &   & 0 & {\Kappa_2}  &  &  &  \\
   &   &   & \ddots  & \ddots &  &  \\
   &   &   &   & 0  & {\Kappa_{d-2}} &  \\
   &   &   &   &   & 0 & {\Kappa_{d-1}} \\
\mbox{\Large $O$} &   &   &   &   &   & 0
\end{array}
\right) \,\,\,.\label{eqn:fblock1}
\eeq
Let us apply the ${A{A^{\ast}}}$-equation in the previous section. The
deformation parameter $t$ coupled to the charge one field
${\ftu{(1)}}$ associated with a {\kae} form $e$ is the coordinate on
the complexified {\kae} moduli space of the ${A}$(M)-model. In our
case, we can take the hermitian metrics
diagonally,
\[
{\gs}:=\mbox{diag}({\,{e^{q_{0}}}\,\,{e^{q_{1}}}\,\cdots \,
{e^{q_{d-1}}}\,\,{e^{q_{d}}}\,})\,\,\,.
\]
Substituting this matrix into the $A{A^{\ast}}$-equation, we obtain
a set of differential equations,
\beq
&&{\del_{\bar{t}}}{\del_t}{q_0}+
{\left|{\Kappa_0}\right|}^{2}  {e^{{q_1}-{q_0}}}=0\,\,\,,\nonumber \\
&&{\del_{\bar{t}}}{\del_t}{q_l}+
{\left|{\Kappa_l}\right|}^{2}  {e^{{q_{l+1}}-{q_l}}}-
{\left|{\Kappa_{l-1}}\right|}^{2}  {e^{{q_{l}}-{q_{l-1}}}}
=0\,\,\,,\,\,\,(l=1,2,\cdots ,d-1) \,\,\,,\nonumber \\
&&{\del_{\bar{t}}}{\del_t}{q_d}-
{\left|{\Kappa_{d-1}}\right|}^{2}  {e^{{q_d}-{q_{d-1}}}}=0\,\,\,,
\label{eqn:todaeq}
\eeq
This set can be rewritten in a compact form by introducing new
variables ${\varphi_l}$,
\beqs
&&{\varphi_l}:={q_l}-{q_{l-1}}+\log \left({{\Kappa_{l-1}}{\bar{\Kappa}_{l-1}}
}\right)\,\,\,\,\,(l=1,2, \cdots ,d)\,\,\,,\\
&&{\del_{\bar{t}}}{\del_t}{\varphi_l}=
{\sum^{d}_{m=1}}{K_{lm}}{e^{\varphi_m}}\,\,\,\,\,(l=1,2, \cdots ,d-1,d)\,\,\,,
\eeqs
where the coefficients ${K_{lm}}$ are given by changing the sign of each
component of the Cartan matrix of the Lie algebra ${A_d}$.
The above system is the A-type Toda equation system. We make one remark here;
The ${A{A^{\ast}}}$-equation system in the ${{CP}^{d+1}}$ model is that of the
affine A-type Toda theory. In our Calabi-Yau $d$-fold case, the system is
 that of non-affine A-type Toda theory. This difference stems from
the property of the first Chern classes ${c_1}$(M)
of the {\kae} manifolds M.
The virtual dimension (the ghost number anomaly) of the correlators in the
A(M)-model can be written,
\beq
({{\dim_{\bc}}M})\cdot (1-g)+{\int_{\Sigma}}
{X^{\ast}}{c_1}(M)\,\,\,,\label{eqn:hozon}
\eeq
where $g$ is a genus of a Riemann surface ${\Sigma}$.
The second term in the above formula
depends on the degree of the map $X$.
For each fixed degree $n$ of the map, there exists one topological
selection rule.
For the ${{CP}^{d+1}}$ case, this second term is
${(d+2)\,n}$.
Also we deform some topological field theory by adding only operators
associated to a {\kae} form. Then the degree of the observables are
conserved in each fusion. For the ${{CP}^{d+1}}$ case, the next fusion
is non-vanishing generally,
\[
\ftu{(1)}\cdot \ftu{(d+1)}\neq 0\,\,\,,
\]
and leads to an affine Toda theory.
On the other hand, the second term vanishes for the manifolds with the
vanishing first Chern classes (Calabi-Yau manifolds).
In these cases, the following fusion leads to zero,
\[
\ftu{(1)}\cdot \ftu{(d)} =0\,\,\,.
\]
It is this reason that the ${A{A^{\ast}}}$-equations for the ${{CP}^{d+1}}$
model is affine Toda, but that for
Calabi-Yau manifolds are non-affine Toda.

Return to the analysis of the Toda equation in our case.
%Following the general analysis about the Toda theories {\ref{LS,Ger}},
We solve the
equation explicitly. Firstly let us introduce a set of Grassmann variables
${\xi^a}\,\,(a=0,1,\cdots ,d)$
and derivatives
${\displaystyle \frac{\del}{\del {\xi^b}}}\,\,(b=0,1,\cdots ,d)$,
which satisfy the anti-commutation relations,
\beqs
&& \ac{\xi^a}{\xi^b}= \ac{\displaystyle \frac{\del}{\del {\xi^a}}}
{\displaystyle \frac{\del}{\del {\xi^b}}}=0\,\,\,,\\
&&\ac{\xi^a}
{\displaystyle \frac{\del}{\del {\xi^b}}}={\delta_{ab}}\,\,\,.
\eeqs
Secondly we define the functions ${\zeta^{+}}$, ${\zeta^{-}}$
by using arbitrary functions ${({\zeta^{+}_{a}},{\zeta^{-}_{b}})}$,
\beq
&&{\zeta^{+}}(t):={\sum^{d}_{a=0}}{\zeta^{+}_{a}}(t)\cdot {\xi^{a}}\,\,\,,\\
&&{\zeta^{-}}({\bar{t}}):=
{\sum^{d}_{b=0}}{\zeta^{-}_{b}}({\bar{t}})\cdot
{\displaystyle \frac{\del}{\del{\xi^{b}} }}\,\,\,,
\eeq
where ${\{{\zeta^{+}_{a}}\}}$ ({\, ${\{{\zeta^{-}_{b}}\}}$\,})
are holomorphic (anti-holomorphic)
functions with respect to the variable $t$.
The set of solutions for ${e^{q_n}}$ is obtained,
\beqs
&&{e^{q_n}}= \frac{{\sfh}_{n+1}}{{\sfh}_{n}}\cdot
{\displaystyle \frac{B_0}{{\left|{{\Kappa_0}{\Kappa_1}\cdots {\Kappa_{n-1}}
}\right|}^2}}\,\,\,,\\
&&{{\sfh}_n}=
\begin{array}[t]{c}
 {\displaystyle \sum }\\
{\scriptstyle  0\leq {i_1} < \cdots < {i_n} \leq d }
\end{array}
{\left|{ {{({\zeta^{+}_{i_1}})}}{{({\zeta^{+}_{i_2}})}}\cdots
{{({\zeta^{+}_{i_n}})}}
}\right|}_{t}
 \times {\left|{ {{({\zeta^{-}_{i_1}})}}{{({\zeta^{-}_{i_2}})}}\cdots
{{({\zeta^{-}_{i_n}})}}
}\right|}_{\bar{t}} \,\,\,,\,\,\,(n=1,2,\cdots )\,\,\,,\\
&&{H_0}:=1\,\,\,,
\eeqs
with one relation,
\beq
1&=& {{({\zeta^{+}})}}{{({\zeta^{+}})}'}{{({\zeta^{+}})}''}\cdots
{{({\zeta^{+}})}^{(d)}}\nonumber \\
&&\times {{({\zeta^{-}})}}{{({\zeta^{-}})}'}{{({\zeta^{-}})}''}\cdots
{{({\zeta^{-}})}^{(d)}}\label{eqn:feq2} \,\,\,,\\
&& \,\,\,\mbox{where}\,\,\,
{{({\zeta^{+}})}^{(n)}}={\del^{n}_{t}}{\zeta^{+}}\,\,\,,\,\,\,
{{({\zeta^{-}})}^{(n)}}={\bar{\del}^{n}_{\bar{t}}}{\zeta^{-}}\,\,\,,\nonumber
\eeq
where we used a notation to save the space,
\beqs
{{\left|{\, {a_1}\,{a_2}\,\cdots \,{a_m}\,
}\right|}_x}
 :=\left|
\begin{array}{cccc}
{a_1} & {a_2} & \cdots & {a_m} \\
{\del_x}{a_1} & {\del_x}{a_2} & \cdots & {\del_x}{a_m} \\
\vdots & \vdots &      & \vdots \\
{\del^{m-1}_x}{a_1} & {\del^{m-1}_x}{a_2} & \cdots & {\del^{m-1}_x}{a_m}
\end{array}
\right| \,\,\,.
\eeqs
Also the ${B_0}$ is some function represented as products of some pure
holomorphic functions and anti-holomorphic ones, which are
determined from the boundary condition.
We use a set of functions as candidate for the general solution
(\ref{eqn:todaeq}),
\beq
&&{\zeta^{+}_{a}}=
\left({ {\vpi^{-1}_{0}}\cdot {\prod^{d-1}_{j=0}}{\Kappa^{-\frac{d-j}{d+1}}_{j}}
}\right)\cdot {\vpi_{a}}\,\,\,,\,\,\,(a=0,1,\cdots ,d) \,\,\,,\\
&&{\zeta^{-}_{b}}=
\left({ {\bar{\vpi}^{-1}_{0}}\cdot
{\prod^{d-1}_{j=0}}{\bar{\Kappa}^{-\frac{d-j}{d+1}}_{j}}
}\right)\cdot {\bar{\vpi}_{d-b}}\cdot {{(-1)}^{d-b}}
\,\,\,,\,\,\,(b=0,1,\cdots ,d) \,\,\,,
\eeq
where the ${\Kappa_j}$ are the three point couplings written in
(\ref{eqn:fblock1}) and the ${\bar{\Kappa}_j}$ are their complex conjugate.
The ${\vpi_a}(z)$ are the solutions of the Picard-Fuchs
equation and written by using
the Schur polynomials,
\beqs
&&{\vpi_a}(z)={\vpi_0}(z)\cdot
{S_a}({{\tilde{x}_1},{\tilde{x}_2},\cdots ,
{\tilde{x}_a}})
\,\,\,,\,\,\,(a=0,1,\cdots ,d)\,\,\,,\\
&&{\hat{\varpi}_0}({z\,;\,\rho}):=
{\sum^{\infty}_{n=0}}
\frac{\displaystyle \Gamma {\!}({N({n+\rho})+1})}
{\displaystyle \Gamma {\!}({N{\rho}+1})}
\cdot
{{\left[{
\frac{\displaystyle \Gamma {\!}({\rho +1})}
{\displaystyle \Gamma {\!}({n+\rho +1})}
}\right]}^{N}}
\cdot
{{({N\psi})}^{-N({n+\rho})}}\,\,\,,\\
&&{\tilde{x}_m}:=\frac{1}{m!}{{\cal D}^{m}_{\rho}}
{\left.{ \log {\hat{\varpi}_0}(z\,;\,{\rho})}\right|}_{\rho =0}\,\,\,,
\,\,\,{{\varpi}_0}({z}):=
{\hat{\varpi}_0}({z\,;\,\rho}){\Big|_{\rho =0}}\,\,\,,\\
&&z:={{({N\psi})}^{-N}}\,\,\,,
\eeqs
Also the ${\bar{\vpi}_a}({\bar{z}})$ are complex conjugate of the
${\vpi_a}(z)$.
These functions satisfy a condition,
\beq
{{\sfh}_{d+1}}= {\left|{ {{({\zeta^{+}_{0}})}}{{({\zeta^{+}_{1}})}}\cdots
{{({\zeta^{+}_{d}})}}
}\right|}_{t}
 \times {\left|{ {{({\zeta^{-}_{0}})}}{{({\zeta^{-}_{1}})}}\cdots
{{({\zeta^{-}_{d}})}}
}\right|}_{\bar{t}} =1 \,\,\,,
\eeq
where we used a relation,
\beqs
&&{{\left|{\,{\omega_0}\,{\omega_1}\,\cdots \,{\omega_d}\,
}\right|}_t}= {{({\omega_0})}^{d+1}}
{{({\Kappa_{d-1}})}^{1}}
{{({\Kappa_{d-2}})}^{2}}
\cdots
{{({\Kappa_{1}})}^{d-1}}
{{({\Kappa_{0}})}^{d}} \,\,\,,\\
&&{\omega_a}:={\displaystyle \frac{\vpi_a}{\vpi_0}}\,\,\,\,(a=0,1,\cdots ,d)
\,\,\,.
\eeqs
In addition, we choose the function ${B_0}$,
\beqs
{B_0}=\left({{\vpi_0}{\bar{\vpi}_0}}\right) \cdot
{\sum^{d-1}_{j=0}}{\left({{\Kappa_j}{\bar{\Kappa}_j}
}\right)}^{-\frac{d-j}{d+1}}\,\,\,.
\eeqs
Under this setup, we can write down solutions
${e^{q_n}}$,
\beqs
%%%%%%%%%%%%%%%%%
{\exp ({q_n})}&=&
{\displaystyle ({{\vpi_0}\cdot {\bar{\vpi}_0}})}\cdot
 {\displaystyle
\frac{
\begin{array}[t]{c}
{\displaystyle \sum} \\
{\scriptstyle 0\leq {i_1} < \cdots < {i_{n+1}} \leq d }
\end{array}
{ {\left|{ \,{\omega_{i_1}}\,{\omega_{i_2}}\,\cdots \,{\omega_{i_{n+1}}}\,
}\right|}_t }\cdot
{ {\left|{ \,{\tilde{\omega}_{i_1}}\,{\tilde{\omega}_{i_2}}\,\cdots
\,{\tilde{\omega}_{i_{n+1}} }\,
}\right|}_{\bar{t}} }}
{
\begin{array}[t]{c}
{\displaystyle \sum} \\
{\scriptstyle 0\leq {j_1} < \cdots < {j_n} \leq d }
\end{array}
{{\left|{ \,{\omega_{j_1}}\,{\omega_{j_2}}\,\cdots \,{\omega_{j_n}}\,
}\right|}_t}\cdot
{{\left|{ \,{\tilde{\omega}_{j_1}}\,{\tilde{\omega}_{j_2}}\,\cdots
\,{\tilde{\omega}_{j_n}}\,
}\right|}_{\bar{t}}}}}
\nonumber \\
&&\times
{\displaystyle
\frac{1}
{{\left|{
{\Kappa_0}{\Kappa_1}\cdots {\Kappa_{n-1}}
}\right|}^2}}
\,\,\,,\\
%%%%%%%%%%%%%%%%%%
&&{\omega_a}:={\displaystyle \frac{\vpi_a}{\vpi_0}}\,\,\,,\,\,\,
{\tilde{\omega}_a}:=
{\displaystyle \frac{\bar{\vpi}_{d-a}}{\bar{\vpi}_0}}\cdot
{({-1})}^{d-a}
\,\,\,.
\eeqs
(These results remind us of the structure of the W-geometry of the
Toda systems \cite{Walg}).
Also the product of the couplings can be expressed as,
\beq
{\displaystyle
{{\left|{
{\Kappa_0}{\Kappa_1}\cdots {\Kappa_{n-1}}
}\right|}^2}}=
{\displaystyle
\frac{
{ {\left|{ \,{\omega_{0}}\,{\omega_{1}}\,\cdots \,{\omega_{n}}\,
}\right|}_t }\cdot
{ {\left|{ \,{\bar{\omega}_{0}}\,{\bar{\omega}_{1}}\,\cdots
\,{\bar{\omega}_{n} }\,
}\right|}_{\bar{t}} }}
{
{{\left|{ \,{\omega_{0}}\,{\omega_{1}}\,\cdots \,{\omega_{n-1}}\,
}\right|}_t}\cdot
{{\left|{ \,{\bar{\omega}_{0}}\,{\bar{\omega}_{1}}\,\cdots
\,{\bar{\omega}_{n-1}}\,
}\right|}_{\bar{t}}}}}\,\,\,.\label{eqn:feq3}
\eeq

Let us make two remarks:
Firstly each summand in the solution ${e^{q_n}}$
behaves in the large radius limit
${\mbox{Im}\,t}\rightarrow \infty $ as,
\beqs
&&{{\left|{\,{\omega_{i_1}}\,\,{\omega_{i_2}}\,\cdots
\,{\omega_{i_{n+1}}}\,
}\right|}_t}\sim {t^{{i_1}+{i_2}+\cdots
+{i_{n+1}}-\frac{1}{2}n({n+1}) }}\,\,\,,\\
&&{{\left|{\,{\tilde{\omega}_{i_1}}\,\,{\tilde{\omega}_{i_2}}\,\cdots
\,{\tilde{\omega}_{i_{n+1}}}\,
}\right|}_{\bar{t}}}\sim {\bar{t}^{({d-{i_1}})+({d-{i_2}})+({d-{i_{n+1}}})
-\frac{1}{2} {n}({n+1}) }}\,\,\,.
\eeqs
On the other hand, the product of the couplings (\ref{eqn:feq3}) tends
to constant classical one in this limit.
When one scales the parameters $({t\,,\,{\bar{t}}})$ as
 ${\lambda \,({t\,,\,{\bar{t}}})}$ in this limit by using a real
positive parameters $\lambda $,
each solution ${e^{q_n}}$ behaves as,
\beqs
{e^{q_n}}\sim {\lambda^{-2n+d}}\,\,\,.
\eeqs
We identify the scale dimension of the $n$-th component of the set of
solutions ${e^{q_n}}$ with $n$
which is the same as that of the hermitian metric
${{g_{n\bar{n}}}=\corr{\ftbu{({\bar{n}})}\Big| \ftu{(n)} }}$.
(The extra scaling exponent
$d$ comes from the axial U(1) anomaly due to the axial couplings
between the U(1) gauge connection ${\sf A}$ and the fermion number current
${\sf J}$).
As a second remark, we consider the rescaling of the periods
${\{{\vpi_a}\}}$. When we rescale ${({\vpi_a},{\bar{\vpi}_b})}$ by
multiplying arbitrary holomorphic (anti-holomorphic) functions
${f(t)}$, ${\bar{f} ({\bar{t}})}$ respectively,
\[
{({\vpi_a},{\bar{\vpi}_b})} \rightarrow
{({f\,\vpi_a},{\bar{f}\,\bar{\vpi}_b})}\,\,\,,
\]
the solutions ${e^{q_n}}$ transform into the forms,
\beqs
&&{e^{q_n}}\rightarrow f(t)\bar{f}({\bar{t}})
{e^{q_n}}\,\,\,,\\
&&\mbox{i.e.}\,\, {q_n}\rightarrow
{q_n}+ \log f(t)+\log \bar{f}({\bar{t}}) \,\,\,.
\eeqs
This means that the {${{q}_n}$}'s are not functions but sections of
holomorphic line bundle ${\cal L}$ over the A-model moduli space.
%\subsection
%{Properties of the Metrics}
%\pr
Next let us investigate properties of the metrics.
Firstly we take the ${e^{q_0}}$.
It is written by a simple calculation,
\beqs
&&{e^{q_0}}=\left({{\vpi_0}{\bar{\vpi}_0}}\right)
\cdot {S_d}({{\tilde{z}_1},{\tilde{z}_2},\cdots ,{\tilde{z}_d}})\,\,\,,\\
&&{\tilde{z}_n}:={\tilde{x}_n}+{{(-1)}^{n}}\cdot
{\overline{\tilde{x}}_n}\,\,\,,\,\,\,
{\tilde{x}_m}:=\frac{1}{m!}{{\cal D}^{m}_{\rho}}
{\left.{\log {\hat{\varpi}_0}({z\,;\,\rho})}\right|}_{\rho =0} \,\,\,,
\eeqs
where the ${S_d}$ is a Schur function.
%(In deriving this result, we use the formula ${{\omega_a}
%={S_d}({{\tilde{x}_1},\cdots ,{\tilde{x}_a}})}$ ).
This corresponds to the two point function of charge zero operators,
\[
{e^{q_0}}={{\gs}_{0\bar{0}}}=\corr{\ftbu{({\bar{0}})}\Big| \ftu{(0)}}\,\,\,.
\]
Generally for restricted {\kae} manifolds, this correlator can be
represented by using the {\kae} potential ${\cal K}$,
\[
{g_{0\bar{0}}}={e^{-{\cal K}}}\,\,\,.
\]
{}From this information, we may put ${{q_0}=-{\cal K}}$. These
have a characteristic property,
\beqs
&&{\cal K}\rightarrow {\cal K}-\log f(t)-\log \bar{f}({\bar{t}})\,\,\,,\\
&& \hspace{1cm}({{\vpi_a},{\bar{\vpi}_b}}) \rightarrow
({f\,{\vpi_a},\bar{f}\,{\bar{\vpi}_b}})\,\,\,.
\eeqs
Several explicit formulae are collected in the Appendix C.
Secondly we study the ${e^{q_1}}$.
{}From the Toda equation ({\ref{eqn:todaeq}}), the ${e^{q_1}}$ is written,
\beqs
{e^{q_1}}&=&{e^{q_0}}\cdot \left({-{\del_{\bar{t}}}
{\del_{{t}}}{q_0} }\right) \\
&=& {e^{q_0}}\cdot \left({{\del_{\bar{t}}}{\del_t}{\cal K} }\right) \,\,\,.
\eeqs
The above derivative of the {\kae} potential
${{\del_{\bar{t}}}{\del_t}{\cal K}}$ is known as
the Zamolodchikov metric ${G_{t\bar{t}}}={\del_{\bar{t}}}{\del_t}{\cal
K}$ in general terms.
A relation between this metric
${G_{t\bar{t}}}$ and the hermitian metric is written as,
\[
{G_{t\bar{t}}}={\displaystyle \frac{{\gs}_{1\bar{1}}}
{{\gs}_{0\bar{0}}}}\,\,\,.
\]
{}From this consideration, we identify ${e^{q_1}}$ with ${{\gs}_{1\bar{1}}}$,
\[
{e^{q_1}}={{\gs}_{1\bar{1}}}=\corr{\ftbu{({\bar{1}})}\Big| \ftu{(1)} }\,\,\,.
\]
We make a conjecture about the interpretation of
the other solutions.
\mbox{}\\
(Conjecture) \\
The solutions ${e^{q_n}}$ of the Toda equation (\ref{eqn:todaeq}) are
identified with the hermitian metrics on the complexified
{\kae} moduli space,
\[
{\displaystyle {\oplus_p}{T^p}({M\,,\,{{\lwedge}^p}
{{\bar{T}}^{\ast}}M})\otimes {\bc}\
}
\]
of the Calabi-Yau A(M) model.\\
In the rest of this section, a genus one partition function is
analyzed by using the results obtained here.

\subsection{The Genus One Partition Function}

\pr
A genus one partition function ${{\sf F}_1}$ is described by the
following new index \cite{BCOV,INDEX},
\beqs
&&{{\sf F}_1}=\frac{1}{2}\cdot {\int_{\cal F}}\frac{{d^2}\tau}{\tau_2}
\tr {{(-1)}^{F}}{F_L}{F_R}{q^{L_0}}{\bar{q}^{\bar{L}_0}}\,\,\,,\\
&&F:={F_L}-{F_R}\,\,,\,\, q:={e^{2\pi i \tau}}\,\,,\,\,
{\bar{q}}:={e^{-2\pi i \bar{\tau}}}\,\,\,,
\eeqs
where the ${F_L}$, ${F_R}$ are left, right fermion number operators.
(Insertion of these operators adjusts the fermion zero modes).
Also the trace in the above formula is over the Ramond sector ground
states and the integral is calculated over the fundamental region
${\cal F}$ of the world sheet torus with a modulus parameter ${\tau =
{\tau_1}+i {\tau_2}}$. This formula is rewritten by the analysis in
the operator formalism,
\beq
{\del_{\bar{\jmath}}}{\del_i}{{\sf F}_1}=\frac{1}{2}
\tr \left\{{ {{(-1)}^{F}}{C_i}{{\bar{C}}_{\bar{\jmath}}} }\right\}
-\frac{1}{24}{G_{i\bar{\jmath}}}\tr {{(-1)}^{F}}\,\,\,,
\eeq
where subscripts ``${i}$'' ``${{\bar{\jmath}}}$'' represent marginal
operators ${\phi_i}$, ${\bar{\phi}_{\bar{\jmath}}}$ on the Ramond
ground states. In our situation, the ``${i}$'' ``${{\bar{\jmath}}}$''
are associated to $A$-model, ${A^{\ast}}$-model operators
${{\phi_A}[{e_i}]}$, ${{\phi_{A^{\ast}}}[{e_j}]}$ constructed from
{\kae} forms ${e_i}$, ${e_j}$ of M respectively.
The symbol ${C_i}$, ${\bar{C}_{\bar{\jmath}}}$ are structure constants
associated to the above marginal operators and the trace is over the
Ramond ground states. In the second term in the right hand side, the
${G_{i\bar{\jmath}}}$ is Zamolodchikov metrics of the Calabi-Yau
{\kae} moduli space. Also the ${\tr {{(-1)}^{F}}}$ is equal to a Euler
number ${\chi}$(M) of the Calabi-Yau d-fold M. The derivatives
${\del_i}$,
${\del_{\bar{\jmath}}}$ in the left hand side are partial derivatives
with respect to marginal coordinates ${t_i}$,
${\bar{t}_{\bar{\jmath}}}$ associated to operators ${{\phi_A}[{e_i}]}$,
${{\phi_{A^{\ast}}}[{e_j}]}$ respectively.

Furthermore the equation can be rewritten as,
\beq
&&{\del_i}{\del_{\bar{\jmath}}}{{\sf F}_1}=\frac{1}{2}
{\del_i}{\del_{\bar{\jmath}}}\log P\,\,\,,\\
&&\log P :={\sum_{p,q}}{{(-1)}^{p-q}}\frac{p+q}{2}
{{\tr}_{p,q}}\left({\log {\gs}}\right)-\frac{{\cal K}}{12}
\tr {{(-1)}^{F}}\,\,\,,
\eeq
where the variables ${p, q}$ in the summand are left, right Ramond
U(1) charges and also equal to the ghost numbers of the $A$-model,
${A^{\ast}}$-model operators. The ``${{\gs}}$'' in the trace are
hermitian metrics ${{\gs}_{p\bar{q}}}=\corr{\ftbu{(\bar{q})}\ftu{(p)}}$
of the {\kae} moduli space. Especially the ${\cal K}$ is a {\kae}
potential of the moduli space and is represented as ${{\cal K}=-\log
{{\gs}_{0\bar{0}}}}$.

\subsection{Application to the Calabi-Yau d-Fold}

\pr
In this subsection, we apply the recipe in the previous subsection to
the Calabi-Yau d-fold ({\ref{eqn:CY}}).
For this Calabi-Yau d-fold $({c=3d})$, the
Ramond U(1) charges ${q_R}$ are written,
\beqs
{q_R}=l-\frac{d}{2}\,\,\,,\,\,\,({l=0,1,\cdots ,d})\,\,\,.
\eeqs
Especially their conformal weights ${h_R}$ is constant,
${{h_R}=\frac{c}{24}=\frac{d}{8}}$.
The corresponding weight ${h_{NS}}$ and U(1) charges ${q_{NS}}$ in the
Neveu-Schwarz sector are obtained by the spectral flow,
\beqs
&&{h_{NS}}={h_R}+\frac{1}{2}{q_R}+\frac{d}{8}=\frac{l}{2}\,\,\,,\\
&&{q_{NS}}={q_R}+\frac{d}{2}=l\,\,\,,\,\,\,({l=0,1,\cdots ,d})\,\,\,.
\eeqs
Note that the factor ${{(-1)}^{p-q}}\cdot {\displaystyle \frac{p+q}{2}}$
 is calculated in
the case ${p=l-\frac{d}{2}}$, ${q=m-\frac{d}{2}}$ as,
\beqs
&&{{(-1)}^{p-q}}\cdot \frac{p+q}{2}=\frac{1}{2}({r-d})\cdot
{{(-1)}^{r}}\,\,\,,\\
&&\,\,\,\,\,\,\,r:=l+m\,\,\,,\,\,\,({l=0,1,\cdots ,d\,;\,
m=0,1,\cdots ,d})\,\,\,.
\eeqs
{}From this consideration, it turns out to be that the primary
horizontal subspace of the de~Rham cohomology of M,
\[
{H^d}(\mbox{M})
:={\displaystyle {\displaystyle \oplus^{d}_{s=0}}}
{H^{d-s,s}}({\mbox{M}})\,\,\,,
\]
does not contribute to the ${{\sf F}_1}$.
Because the Hodge numbers are given as ({\ref{eqn:hodge}}) in our model
({\ref{eqn:CY}}),
 ${\log P}$
is obtained,
\beq
\log P=-\frac{\chi}{12}\cdot {\cal K}+{\sum^{d}_{s=0}}
\left({s-\frac{d}{2}}\right)\cdot \log ({{\gs}_{s\bar{s}}})\,\,\,.
\eeq
By using a reality condition of the hermitian metrics,
\beq
{{\gs}_{d-s, \overline{d-s}}}=\frac{1}{{\gs}_{s\bar{s}}}\,\,\,,
\eeq
we obtain a formula,
\beq
&&\log P=\left({
-\frac{\chi}{12}+\gauss{\frac{d+1}{2}}\gauss{\frac{d+2}{2}}
}\right){\cal K}-{\sum^{\gauss{\frac{d-1}{2}}}_{s=1}}({d-2s})
\log {G_{s\bar{s}}}\,\,\,,\\
&&{G_{s\bar{s}}}:=\frac{{\gs}_{s\bar{s}}}{{\gs}_{0\bar{0}}}\,\,\,,\,\,\,
({s=0,1,\cdots ,d})\,\,\,.
\eeq
We define a symbol ${\gauss{x}}$ as a unique maximal integer which is
not greater than ${x}$.
Thus the partition function ${{\sf F}_1}$ is written as,
\beq
{{\sf F}_1}=\frac{1}{2}\cdot \log
\left[{\exp \left\{{
\left({
-\frac{\chi}{12}+\gauss{\frac{d+1}{2}}\gauss{\frac{d+2}{2}}
}\right){\cal K}
}\right\}
\times {\displaystyle {\prod^{\gauss{\frac{d-1}{2}}}_{s=1}}}
{{\left({G_{s\bar{s}}}\right)}^{-(d-2s)}}\cdot {{|f|}^2}
}\right]\,\,\,,
\eeq
where $f$ is an unknown holomorphic function determined by a boundary
condition. Now we use the mirror symmetry of the d-fold M and its
partner W. Recall that W is constructed as a orbifold from M divided
by the maximally invariant discrete group ${{({\bz}_{d+2})}^{d+1}}$.
The complex moduli space of W has singularities at points ${\psi =0,
{\tilde{\alpha}^a}, \mbox{and } \infty}$ ${({a=0,1,\cdots , d+1\,;\,
\tilde{\alpha} :={e^{\frac{2\pi i}{d+2}}}})}$. Because the behaviour of the
function $f$ is controlled by these singular points, we make a
conjecture about $f$,
\beq
f={\psi^{\alpha}}{{({1-{\psi^{d+2}}})}^{\beta}}\,\,\,,
\eeq
with some constant numbers ${\alpha}$ and ${\beta}$.
Thus we write a genus one partition function of W as,
\beq
&&{{\sf F}_1}({\psi})=\frac{1}{2}\cdot \log
\Biggl[ \exp \left\{{
\left({
-\frac{\chi}{12}+\gauss{\frac{d+1}{2}}\gauss{\frac{d+2}{2}}
}\right){\cal K}
}\right\}\nonumber \\
&& \hspace{11mm} \times {\displaystyle {\prod^{\gauss{\frac{d-1}{2}}}_{s=1}}}
{{\left({ {G_{s\bar{s}}}({\psi}) }\right)}^{-(d-2s)}}\cdot
{{\left|{
{\psi^{\alpha}}{{({1-{\psi^{d+2}}})}^{\beta}}
}\right|}^2}
\Biggr]\,\,\,,\\
&&{\cal K}=-\log {{\gs}_{0\bar{0}}}\,\,\,,\nonumber \\
&&{{\gs}_{0\bar{0}}}=
\left({{\varpi_0}{\bar{\varpi}_0}}\right)\cdot
{S_d}({{\tilde{z}_1}\,,\,{\tilde{z}_2}\,,\,\cdots \,,\,
{\tilde{z}_d}
})\,\,\,,\\
&&{G_{n\bar{n}}}({\psi})={e^{{q_n}-{q_0}}}\nonumber \\
&&=
{\displaystyle \frac{1}{{S_d}({\tilde{z}})}}\cdot
 {\displaystyle
\frac{
\begin{array}[t]{c}
{\displaystyle \sum} \\
{\scriptstyle 0\leq {i_1} < \cdots < {i_{n+1}} \leq d }
\end{array}
{ {\left|{ \,{\omega_{i_1}}\,{\omega_{i_2}}\,\cdots \,{\omega_{i_{n+1}}}\,
}\right|}_{\psi} }\cdot
{ {\left|{ \,{\tilde{\omega}_{i_1}}\,{\tilde{\omega}_{i_2}}\,\cdots
\,{\tilde{\omega}_{i_{n+1}} }\,
}\right|}_{\bar{\psi}} }}
{
\begin{array}[t]{c}
{\displaystyle \sum} \\
{\scriptstyle 0\leq {j_1} < \cdots < {j_n} \leq d }
\end{array}
{{\left|{ \,{\omega_{j_1}}\,{\omega_{j_2}}\,\cdots \,{\omega_{j_n}}\,
}\right|}_{\psi}}\cdot
{{\left|{ \,{\tilde{\omega}_{j_1}}\,{\tilde{\omega}_{j_2}}\,\cdots
\,{\tilde{\omega}_{j_n}}\,
}\right|}_{\bar{\psi}}}}}
\nonumber \\
&&\hspace{11mm} \times
{\displaystyle
\frac{1}
{{\left|{
{\Kappa_0}{\Kappa_1}\cdots {\Kappa_{n-1}}
}\right|}^2}}
\,\,\,.\label{eqn:fzamo}
\eeq
Because the partition function is a zero point function, the
corresponding one ${{\sf F}_1}(t)$ in the $A$(M)-model can be obtained
only by changing parameters from ${\psi}$ to
${t({\psi}):={\tilde{x}_1}}$,
\beq
{{\sf F}_1}(t)={{\sf F}_1}({\psi (t)})\,\,\,.
\eeq
All we have to do is to determine the unknown constant numbers ${\alpha}$
and ${\beta}$. To achieve this purpose, we examine asymptotic behaviours
of the ${{\sf F}_1}$ in some limits. \\
Firstly in the large radius limit, the ${{\sf F}_1}$ becomes,
\beq
{{\sf F}_1}{\Big|_{({t,\bar{t}})\rightarrow \infty }}&=&
\frac{{-1}}{24}({t+\bar{t}}){\int_M}e\wedge {c_{d-1}}\,\,\,
\nonumber \\
&=&({t+\bar{t}})\cdot \frac{{-1}}{24}\cdot
\left\{{
\frac{1}{N^2}\left[{
1-{{({1-N})}^{N}}
}\right]+\frac{1}{2}({N-2})({N+1})
}\right\}\,\,\,,\\
 N{\!}&{:=}&{\!}d+2\,\,\,.\nonumber
\eeq
Secondly we take an asymmetrical limit ${\bar{t}\rightarrow \infty}$
while keeping $t$ fixed. In this limit, the normalized hermitian
metrics ${G_{n\bar{n}}}$ behaves as,
\beq
\log {G_{n\bar{n}}}(t){\!}&{\sim}&{\!}
-\log ({S_d})-\log {{\left|{
{\Kappa_0}{\Kappa_1}\cdots {\Kappa_{n-1}}
}\right|}^2}\nonumber \\
&&+
\log {{\left|{
{\Kappa_0}{\Kappa_1}\cdots {\Kappa_{n-1}}
}\right|}}
+\log \frac{
{\left|{
{\tilde{\omega}_0}{\tilde{\omega}_1}\cdots {\tilde{\omega}_{n}}
}\right|}_{\bar{t}}}
{
{\left|{
{\tilde{\omega}_0}{\tilde{\omega}_1}\cdots {\tilde{\omega}_{n-1}}
}\right|}_{\bar{t}}}\nonumber \\
&&=-\log ({S_d})+({\mbox{anti-holomorphic parts}})+{\cal
O}({\bar{t}^{-1}})\,\,\,.
\eeq
Now the Schur polynomial ${S_d}$ is written down,
\beqs
&&{S_d}=\frac{1}{d!}\cdot {{({t-\bar{t}})}^{d}}+
\frac{1}{(d-2)!}\cdot {{(t-\bar{t})}^{d-2}}\cdot
\left({
{D_{\rho}}t+\overline{{D_{\rho}}t}}\right)
{\Big|_{\rho=0}}+\cdots\,\,\,,\\
&&\,\,\,\,\,\mbox{with }{D_{\rho}}t:=
{{\left({\frac{1}{2\pi i}\cdot
\frac{\del}{\del \rho}}\right)}^2}\log {\hat{\varpi}_0}({z\,;\,\rho})\,\,\,,
\eeqs
and then,
\beqs
\log {S_d}\sim \log {({\bar{t}})^d}+{\cal
O}({\bar{t}}^{-1})\,\,\,,\,\,\,
({\bar{t}\rightarrow \infty})\,\,\,.
\eeqs
Thus the metric is reduced to an asymptotic form,
\beq
\log {G_{n\bar{n}}}(t)\sim ({\mbox{anti-holomorphic parts}})+{\cal
O}({{\bar{t}}^{-1}})\,\,\,.
\eeq
Similarly the {\kae} potential ${\cal K}$ behaves as,
\beqs
{\cal K}&=&-\log {{\gs}_{0\bar{0}}}\\
&\sim & -\log {\varpi_0}+({\mbox{anti-holomorphic parts}})+
{\cal O}({{\bar{t}}^{-1}})\,\,\,,
\eeqs
in the limit ${\bar{t}\rightarrow \infty}$.
By using a transformation property,
\beq
{G_{{n}\bar{n}}}({\psi})=
{{\left({
\frac{\del t}{\del \psi}
}\right)}^n}
{{\left({\overline
\frac{\del t}{\del \psi}
}\right)}^n}\cdot
{G_{n\bar{n}}}({t})\,\,\,,
\eeq
we obtain an equation,
\beq
&&{{\sf F}_1}({\bar{t}\rightarrow \infty})=\frac{1}{2}\cdot
\log \left[{
{{({\varpi_0})}^{-v}}
{{\left({\frac{\del \psi}{\del t}}\right)}^{u}}\cdot {\psi^{\alpha}}
{{({1-{\psi^N}})}^{\beta}}
}\right]\,\,\,,\label{eqn:akotae}\\
&&\chi =\frac{1}{N}\cdot
\left[{
{{(1-N)}^{N}}-1+{N^2}
}\right]\,\,\,,\,\,\,({\mbox{Euler number of M}})\,\,\,,
\label{eqn:bkotae}\\
&&v:=-\frac{\chi}{12}+\gauss{\frac{d+1}{2}}\gauss{\frac{d+2}{2}}\,\,\,,
\label{eqn:ckotae}\\
&&u:=
{\displaystyle
\left\{
\begin{array}{lcl}
\frac{1}{6}\gauss{\frac{d-1}{2}}\gauss{\frac{d+1}{2}}\cdot d & &
({d\,;\,\mbox{odd}}) \\
\frac{1}{6}\gauss{\frac{d-1}{2}}\gauss{\frac{d+1}{2}}\cdot (d+2) & &
({d\,;\,\mbox{even}})
\end{array}
\right.
} \,\,\,.\label{eqn:dkotae}
\eeq
By postulating the regularity at the singular points ${\psi =0}$ and
${\infty}$ of the complex moduli space of W, we obtain the constants
${\alpha}$ and ${\beta}$,
\beq
&&{\alpha}=v\,\,,\,\,\beta =\frac{1}{12}\cdot {N_{d-1}}
-\frac{u+v}{N}\,\,\,,\label{eqn:ekotae}\\
&&{N_{d-1}}=\frac{1}{N^2}\cdot
\left[{
1-{{(1-N)}^{N}}
}\right]+\frac{1}{2}\cdot ({N-2})({N+1})\,\,\,.\label{eqn:fkotae}
\eeq
(The derivation of these are summarized in the Appendix D).
We write the final result of the genus
one partition function ${{\sf F}_{1}}$
of the Calabi-Yau $d$-fold embedded in the
projective space ${{CP}^{d+1}}$,
\beq
&&{{\sf F}_{1}}=\frac{1}{2}\cdot \log
\left[{ {e^{{v}\cdot {\cal K}}}
\cdot \left({
{\displaystyle \prod^{\gauss{\frac{d-1}{2}}}_{n=1}}
{\left({G_{n\bar{n}} }\right)}^{-(d-2\,n)}
}\right)
\times
 {\left|{ {\psi^{\alpha}} {{({1-{\psi^N}})}^{\beta}}    }\right|}^2
}\right] \,\,\,,\label{eqn:gkotae}
%&&N:=d+2 \,\,\,,\\
%&& {v}:= \left[{\frac{d+1}{2}}\right] \cdot
%\left[{\frac{d+2}{2}}\right] -\frac{\chi}{12}\,\,\,,\\
%&&{u}:=\left\{
%\begin{array}{lcl}
%{ \frac{1}{6} \left[{\frac{d+1}{2}}\right] \cdot
%\left[{\frac{d-1}{2}}\right] \cdot (d+2)
%} & & ({d\,;\,\mbox{even}}) \\
%{ \frac{1}{6} \left[{\frac{d+1}{2}}\right] \cdot
%\left[{\frac{d-1}{2}}\right] \cdot d
%} & & ({d\,;\,\mbox{odd}})
%\end{array}
%\right.\,\,\,,\\
%&&{\alpha}:={v}\,\,\,,\\
%&&{\beta}:=\frac{{(-1)}^{d-1}}{12}\,{N_{d-1}}-\frac{{u}+{v}}{N} \,\,\,,\\
%&&{\chi}:=\frac{1}{N}\cdot \left[{{(1-N)}^N -1+{N^2} }\right] \,\,\,,\\
%&&{N_{d-1}}:=%{\int_M} J\wedge {c_{d-1}}(M)=
%\frac{1}{N^2}\left[{1-{{(1-N)}^N} }\right]
%+\frac{1}{2}{(N-2)}{(N+1)}\,\,\,.
\eeq
with several parameters ({\ref{eqn:bkotae}})-({\ref{eqn:fkotae}}).
Especially when the parameter $\bar{t}$ tends to infinity with the
parameter t fixed
, the above ${{\sf F}_1}$ ({\ref{eqn:gkotae}})
turns into a formula,
\beq
{{\sf F}_{1}}({\bar{t}\rightarrow \infty})=\frac{1}{2}\cdot
\log \left[{ {\left({\frac{\psi}{\varpi_0} }\right)}^{{\!}v}
\cdot {({1-{\psi^N}})}^{\beta} \cdot
{\left({\frac{\del \psi}{\del t} }\right)}^{{\!}u}
%{\displaystyle
%\frac{1}
%{
%{\Kappa_0}{\Kappa_1}\cdots {\Kappa_{n-1}}
%}}
}\right] \,\,\,.\label{eqn:hkotae}
\eeq
In the next section, we explain the meaning of this formula.

\section{Interpretation of the Result}

\pr
In this section, we explain the meaning of the ${{\sf F}_1}$ obtained
in the previous section from the point of view of the topological
field theory.

\subsection{The Asymmetrical Limit}

\pr
In order to consider the geometrical meaning of the ${{\sf F}_1}$, we
consider the limiting case ${\bar{t}\rightarrow \infty }$. The bosonic
part ${S_0}$ of the action ({\ref{eqn:connect}}) can be written,
\beqs
{S_0}&=& t{\int_{\Sigma}}{d^2}z
\, {g_{i\bar{\jmath}}}\, {\del_z}{X^i}{\del_{\bar{z}}}{X^{\bar{\jmath}}}
+\bar{t}{\int_{\Sigma}}{d^2}z\,
{g_{i\bar{\jmath}}}\, {\del_{\bar{z}}}{X^i}{\del_z}{X^{\bar{\jmath}}}\,\,\,,
\eeqs
with appropriate constant shifts of the parameters
${(t\,,\,{\bar{t}})}$.
Obviously the dominant configuration of the boson field in the
asymmetrical limit ({${\bar{t}\rightarrow \infty}$ with $t$ fixed}) is
the holomorphic mapping from ${\Sigma}$ to M,
\[
{\del_{\bar{z}}}{X^i}=0\,\,\,.
\]
On the other hand, the ${S_0}$ is expressed as,
\beqs
{S_0}&=& \frac{t+\bar{t}}{2}\cdot {\int_{\Sigma}}{d^2}z\,
{g_{i\bar{\jmath}}}\left({
{\del_z}{X^i}{\del_{\bar{z}}}{X^{\bar{\jmath}}}+
{\del_{\bar{z}}}{X^i}{\del_z}{X^{\bar{\jmath}}}
}\right)+
\frac{t-\bar{t}}{2}\cdot {\int_{\Sigma}}{d^2}z\,
{g_{i\bar{\jmath}}}\left({
{\del_z}{X^i}{\del_{\bar{z}}}{X^{\bar{\jmath}}}-
{\del_{\bar{z}}}{X^i}{\del_z}{X^{\bar{\jmath}}}
}\right)\,\,\,.
\eeqs
The second term is seen to be an integral of the pulled-back {\kae}
form $e$ of M to $\Sigma $ and can be rewritten,
\beqs
&& %\frac{1}{2}\,
{\int_{\Sigma}}{d^2}z\,
{g_{i\bar{\jmath}}}\left({
{\del_z}{X^i}{\del_{\bar{z}}}{X^{\bar{\jmath}}}-
{\del_{\bar{z}}}{X^i}{\del_z}{X^{\bar{\jmath}}}
}\right)
=-\sqrt{-1}\,{\int_{\Sigma}}{X^{\ast}}(e)\,\,\,,\\
&&e:=%\frac{1}{2}\,
\sqrt{-1}\,{g_{i\bar{\jmath}}}\,d{X^i}\wedge d{X^{\bar{\jmath}}}\,\,\,.
\eeqs
(In particular, it is independent of the complex structure of the
world sheet ${\Sigma}$). With an appropriate normalization, this gives
a degree of the map ${X}$. Then the path integral of the bosonic part
is reduced to the integrals of the holomorphic maps classified by
their definite degrees.
Now recall the term coupled with the parameter ${\bar{t}}$ in
({\ref{eqn:connect}}),
\beq
\bar{t}{\int_{\Sigma}} \ac{\widetilde{Q}_R}
{\cm{Q_L}{\sqrt{-1}\,{g_{i\bar{\jmath}}}\,{\psi^{i}_{R}}
{\psi^{\bar{\jmath}}_{L}} }}\,\,\,. \label{eqn:connect2}
\eeq
In the limit ${\bar{t}\rightarrow \infty }$, this term should be
decoupled in order to give non-vanishing contributions to the ${{\sf
F}_1}$ in carrying out the path integration. The natural recipe to
decouple this term from the theory is to associate the operators
${\widetilde{Q}_R}$, ${Q_L}$ to some BRST operator. In that situation,
the above term works trivially and does not contribute in the physical
situation. When we combine these operators into one operator
${{Q^{(+)}}=
{\widetilde{Q}_R}+{Q_L} }$, the theory turns back to the $A$(M)-model.
We make a remark; The meaning of the holomorphicity varies when one
deforms the complex structure of the Riemann surface ${\Sigma}$.
Because the integral of the world sheet moduli is performed in the
genus one case, we should take into account of the variation of the
complex structure of the world sheet (i.e. the topological gravity).
In the next subsection, we study a Calabi-Yau $A$(M)-model coupled to
the two-dimensional topological gravity.

\subsection{The $A$-Model Coupled to the Topological Gravity}

\pr
We consider a topological gravity system. However we take an attention
to the complex moduli of the world sheet mainly and omit the
diffeomorphism parts and fix the local Weyl scaling eventually.
The action $S$ in this system consists of two parts; the topological
gravity part ${S_G}$ and the $A$(M)-matter part ${S_M}$ coupled with
the gravity. Firstly we investigate the matter part ${S_M}$
\cite{WIT,hori},
\beq
{S_M}&=&{\int_{\Sigma}}{d^2}z
\Biggl[
{g_{i\bar{\jmath}}}\,{\del_{\bar{z}}}{X^i}{\del_z}{X^{\bar{\jmath}}}
+\sqrt{-1}{\rho_{z i}}
\left({{D_{\bar{z}}}{\chi^i}
+{\chi^{z}_{\bar{z}}}{\del_z}{X^i}
}\right)\nonumber \\
&& +\sqrt{-1}{\rho_{\bar{z} \bar{\imath}}}
\left({{D_{z}}{\chi^{\bar{\imath}}}+{\chi^{\bar{z}}_{z}}
{\del_{\bar{z}}}{X^{\bar{\imath}}}
}\right)
-{\chi^k}{\chi^{\bar{l}}}{{R^{i}_{j}}_{k\bar{l}}}
{\rho_{zi}}{\rho^{j}_{\bar{z}}}+{\chi^{z}_{\bar{z}}}
{\chi_{z}^{\bar{z}}}{\rho_{zi}}{\rho^{i}_{\bar{z}}}
\Biggr]\,\,\,,\label{eqn:matter1}
\eeq
where the fields ${\chi^{z}_{\bar{z}}}$, ${\chi_{z}^{\bar{z}}}$ are
superpartners of the complex structure ${J^{\mu}_{\nu}}$ on the
${\Sigma}$ and its conjugate. The BRST transformations of these fields
are collected as,
\beqs
&& \delta {X^i}={\chi^i}\,\,\,,\,\,\,\delta {\chi^i}=0\,\,\,,\\
&& \delta {X^{\bar{\imath}}}={\chi^{\bar{\imath}}}
\,\,\,,\,\,\,\delta {\chi^{\bar{\imath}}}=0\,\,\,,\\
&& \delta {\rho^{\bar{\imath}}_{z}}=\sqrt{-1}{\del_z}{X^{\bar{\imath}}}
-{\Gamma^{\bar{\imath}}_{\bar{\jmath}\bar{k}}}{\chi^{\bar{\jmath}}}
{\rho^{\bar{k}}_{z}}\,\,\,,\\
&& \delta {\rho_{\bar{z}}^{i}}=\sqrt{-1}{\del_{\bar{z}}}{X^{i}}
-{\Gamma^{i}_{j k}}{\chi^{j}}
{\rho^{k}_{\bar{z}}}\,\,\,,\\
&& \delta {J^{z}_{\bar{z}}}=-2 \sqrt{-1}
{\chi^{z}_{\bar{z}}}\,\,\,,\,\,\,\delta
{\chi^{z}_{\bar{z}}}=0\,\,\,,\\
&& \delta {J_{z}^{\bar{z}}}= 2 \sqrt{-1}
{\chi_{z}^{\bar{z}}}\,\,\,,\,\,\,\delta
{\chi_{z}^{\bar{z}}}=0\,\,\,.
\eeqs
This system is quasi-topological and the weak coupling limit gives
exact results. In fact, the path integral is dominated by the
holomorphic instanton configurations, which are defined in the space
${{{\cal C}_{\Sigma}}\times \mbox{Map} ({\Sigma\,,\,\mbox{M}})}$ as,
\beqs
{Inst} :=\left\{{
({J\,,\,f})\,;\,\,f:{\Sigma_J}\rightarrow \mbox{M} \,\,\,\,\,
({\mbox{holomorphic}})
}\right\}\,\,\,.
\eeqs
The  complex structure ${{\cal C}_{\Sigma}}$ of the Riemann surface
${\Sigma}$ is described by the $(1,1)$ tensor $J$ on ${\Sigma}$.
(We use the notation in \cite{hori}). The holomorphic property of the
map $f$ is written as,
\beqs
\frac{1}{2}d{x^{\mu}}({{\delta^{\nu}_{\mu}}+\sqrt{-1}{J^{\nu}_{\mu}}
}){\del_{\nu}}{f^i}=0\,\,\,,
\eeqs
where the $\left\{{x^{\mu}}\right\}$ is the local real
coordinate on ${\Sigma}$.

Now let us consider an infinitesimal deformation of the configuration
${({J\,,\,f})\rightarrow ({{J+\delta J}\,,\,{f+\delta f}})}$.
The deformed configuration remains in the space $Inst$ if the
following equation is satisfied,
\beqs
{\del_{\bar{z}}}(\delta {f^i})+\frac{\sqrt{-1}}{2}
\delta {J^{z}_{\bar{z}}}{\del_z}{f^i}=0\,\,\,.
\eeqs
The tangential direction of the $Inst$ is controlled by the ghost zero
modes ${({\chi^i}\,,\,{\chi^{z}_{\bar{z}}})}$ with the relation,
\beq
{\del_{\bar{z}}}{\chi^i}+{\chi^{z}_{\bar{z}}}{\del_z}{f^i}=0\,\,\,.
\label{eqn:tangen}
\eeq
The moduli space
%${\overline{\cal M}_{g,s}}$
of the genus $g$ Riemann
surface with $s$-distinct punctures is described by the fermionic
fields ${\chi^{z}_{\bar{z}}}$ and ${\chi^{\bar{z}}_{z}}$. On the other
hand, the transversal direction to the $Inst$ is represented by the
pair $({v^i}\,,\,{\omega})$. The ${v^i}$ is the variation of the map
${f^i}$  $({\delta {f^i}={v^i}})$ and belongs to the space
${({H^0}({f^{\ast}}T\mbox{M}))^{\bot}}$. The $\omega $ is a Beltrami
differential with properties,
\beqs
&&\delta {J^{z}_{\bar{z}}}=-2\sqrt{-1}\, {\omega^{z}_{\bar{z}}}\,\,\,,\\
&&{\omega^{z}_{\bar{z}}}{\del_z}{f^i}\neq 0\,\,\,.
\eeqs
Also their complex conjugate $({{\bar{v}}\,,\,{\bar{\omega}}})$
satisfies a condition,
\beqs
\bar{v}\in {({H^0}{({f^{\ast}}{T^{\ast}}M)})^{\bot}}\,\,\,,
\eeqs
and the ${\bar{\omega}}$ can be expressed by a holomorphic map ${f^i}$
and an anti-ghost ${\rho_{zi}}$ as,
\beqs
{g_{\bar{z}z}}{\bar{\omega}^{\bar{z}}_{z}}={\rho_{zi}}{\del_z}{f^i}\,\,\,.
\eeqs
When one deforms the pair $({J\,,\,f})$ infinitesimally into
$({{J+\delta J}\,,\,{f+\delta f}})$, the action also changes. The
quadratic parts with respect to the variation fields can be
obtained,
\beqs
{S^{(2)}}&=& {S_1}+{S_2}\,\,\,,\\
{S_1}&:=& {\int_{\Sigma}}{d^2}z \Biggl\{
\sqrt{-1}{\rho_{zi}}
\left({{\del_{\bar{z}}}{\chi^i}+{\chi^{z}_{\bar{z}}}{\del_z}{f^i}
}\right)+
\sqrt{-1}{\rho_{\bar{z}\bar{\imath}}}
\left({{\del_{z}}{\chi^{\bar{\imath}}}
+{\chi_{z}^{\bar{z}}}{\del_{\bar{z}}}{f^{\bar{\imath}}}
}\right)\\
&& -{{R^{i}_{j}}_{k\bar{l}}}{\rho_{zi}}{\rho^{j}_{\bar{z}}}
{\chi^{k}}{\chi^{\bar{l}}}+{\rho_{zi}}{\rho^{i}_{\bar{z}}}{\chi^{z}_{\bar{z}}}
{\chi_{z}^{\bar{z}}}
\Biggr\}\,\,\,,\\
{S_2}&=& {\int_{\Sigma}}{d^2}z \Biggl\{
-{g^{\bar{\imath}j}}{v_{\bar{\imath}}}{D_{\bar{z}}}{D_{z}}{\bar{v}_j}
+{D_{\bar{z}}}{v_{\bar{\imath}}}\cdot {\bar{\omega}^{\bar{z}}_{z}}
{\del_{\bar{z}}}{f^{\bar{\imath}}}
+{\omega^{z}_{\bar{z}}}{\del_z}{f^i}\cdot {D_z}{\bar{v}_i}\\
&& +{g_{i\bar{\jmath}}}\,
{\omega^{z}_{\bar{z}}}{\del_z}{f^i}\cdot
{\bar{\omega}^{\bar{z}}_{z}}{\del_{\bar{z}}}{f^{\bar{\jmath}}}
 -\sqrt{-1}{v^i}{\tilde{\varphi}_{\bar{z}zi}}+\sqrt{-1}
{\overline{\tilde{\varphi}}_{z\bar{z}i}}{\bar{v}^{\bar{\imath}}}\\
&& -\sqrt{-1}{\omega^{z}_{\bar{z}}}{D_z}{\chi^i}\cdot {\rho_{zi}}+
\sqrt{-1}{\rho_{\bar{z}\bar{\imath}}}{D_{\bar{z}}}{\chi^{\bar{\imath}}}
\cdot {\bar{\omega}^{\bar{z}}_{z}}
\Biggr\}\,\,\,,\\
&& {\tilde{\varphi}_{\bar{z}zi}}:=
{\chi^k}{\del_{\bar{z}}}{f^{\bar{l}}}\cdot {{R^{j}_{i}}_{k\bar{l}}}
{\rho_{zj}}-{D_z}{{({\chi^{z}_{\bar{z}}}{\rho_z})}_i}\,\,\,,\\
&& {\overline{\tilde{\varphi}}_{z\bar{z}\bar{\imath}}}:=
{\rho_{\bar{z}\bar{\jmath}}}{\chi^{\bar{k}}}
{\del_{z}}{f^{l}}\cdot {{R^{\bar{\jmath}}_{\bar{\imath}}}_{\bar{k} l }}
-{D_{\bar{z}}}
{{( {\rho_{\bar{z}}}{\chi^{\bar{z}}_{z}} )}_{\bar{\imath}}}\,\,\,.
\eeqs
The ghosts are decomposed into sums of the tangential parts and the
transversal parts of the $Inst$. When one carries out the path
integrals in the transversal parts, the contributions from the bosonic
parts and the fermionic parts cancel with each other. Then one obtains
an effective action of the zero mode components,
\beq
{S_{eff}}&=& {\int_{\Sigma}}
\Biggl\{
-{\chi^{k}}{\chi^{\bar{l}}}{{R^{i}_{j}}_{k\bar{l}}}
{\rho_{zi}}{\rho^{j}_{\bar{z}}}+{\chi^{z}_{\bar{z}}}
{\chi_{z}^{\bar{z}}}{\rho_{zi}}{\rho^{i}_{\bar{z}}}
+{{\overline{\tilde{\varphi}}_{z\bar{z}}}^i}
{{\left\{{G(\tilde{\varphi})}\right\}}_i}\nonumber \\
&& -\left({
\bar{\rho}
({\bar{D}\bar{\chi}})-\overline{\tilde{\varphi}}G\bar{D}{\bar{f}_{\sharp}}
}\right)
{\left({
{{{{{}^{t}}{f_{\sharp}}}({1-DG\bar{D}}){\bar{f}_{\sharp}}}}}\right)^{-1}}
\left({
{(D\chi )\rho +{{}^{t}}{f_{\sharp}}DG({\tilde{\varphi}})}
}\right)
\Biggr\}\,\,\,,\label{eqn:effect}
\eeq
where the fermions
$({{\chi^i}\,,\,{\rho^{i}_{\bar{z}}}\,,\,{\rho_{zi}}})$ satisfy the
zero mode conditions,
\beqs
&&{\del_{\bar{z}}}{\chi^i}+{\chi^{z}_{\bar{z}}}{\del_z}{f^i}=0\,\,\,,\\
&&{\del_{\bar{z}}}{\rho_{zi}}=0\,\,\,,\,\,\,{\rho_{zi}}{\del_z}{f^i}=0\,\,\,.
\eeqs
Also the symbol ``$G$'' means a Green's function of the operator
${\bar{D}D}$ and the mappings ${f_{\sharp}}$ and ${\bar{f}_{\sharp}}$
are defined on the ${\omega}$ and ${\bar{\omega}}$ respectively,
\beqs
&&{{({{f_{\sharp}}\omega})}_{\bar{z}\bar{\imath}}}:=
{g_{i\bar{\jmath}}}\,{\del_z}{f^j}\cdot {\omega^{z}_{\bar{z}}}\,\,\,,\\
&&{{({{\bar{f}_{\sharp}}\bar{\omega}})}_{{z}{i}}}:=
{g_{i\bar{\jmath}}}\,{\del_{\bar{z}}}{f^{\bar{\jmath}}}\cdot
{\bar{\omega}_{z}^{\bar{z}}}\,\,\,.
\eeqs
In the next subsection, we give a geometrical meaning of the effective
action (\ref{eqn:effect}) for the (anti)-ghost zero modes.

\subsection{The Geometrical Meaning}

\pr
In the last subsection, we obtain an effective action for the
(anti)-ghost
% $({{\rho}\,,\,{\chi}})$
zero modes.
If one integrates this with respect to the (anti)-ghost zero modes,
the result gives a top Chern class of the vector bundle {\Large
${\nu}$} over the holomorphic instanton space $Inst$. The fiber of the
bundle {\NU} is spanned by the anti-ghost zero modes,
\beqs
{\del_{\bar{z}}}{\rho_{zi}}=0\,\,\,,\,\,\,
{\rho_{zi}}{\del_z}{f^i}=0\,\,\,.
\eeqs
On the bundle {\NU},
a covariant derivative ${\nabla}$
is defined as
\cite{hori},
\beqs
\nabla\rho &=& \Bigl\{
\delta {\rho_i}-\delta {f^k} {\Gamma^{j}_{ki}}{\rho_j}
+\frac{\sqrt{-1}}{2}\cdot d\bar{z}\,\delta {J^{z}_{\bar{z}}}
{\rho_{zi}}\\
&& -dz\,{D_z}{{(G({\tilde{\varphi}}))}_i}+
dz\,{{(df)}^{-1}}
\left\{{
({D{\delta^{1,0}}f})\rho +{{}^{t}}{f_{\sharp}}DG({\tilde{\varphi}})
}\right\}
\Bigr\}\otimes d{X^i} {{\Bigg|}_{(J,f)}}\,\,\,,
\eeqs
where ${{(df)}^{-1}}$ is a Green's function of the operator $df$,
\[
df\,;\,\,{\xi_{zi}}\mapsto {\xi_{zi}}{\del_z}{f^i}\,\,\,.
\]
The second and the third terms originate from the variations of the
map ${f+\delta f}$ and the complex structure ${J+\delta J}$
respectively. The presence of the fourth and the fifth terms
guarantees the holomorphicity and the conormal property of the
${\rho}$ with respect to the $({{J+\delta J}\,,\,{f+\delta f}})$.
By using this connection on {\NU}, the effective action can be
rewritten as,
\beqs
{S_{eff}}=-\left({\rho \,,\, \cm{{\nabla}^{0,1}}{{\nabla}^{1,0}}
\rho}\right)\,\,\,,
\eeqs
with ${\delta {f^i}={\chi^i}}$, ${\delta {J^{z}_{\bar{z}}}=
-2\sqrt{-1} {\chi^{z}_{\bar{z}}} }$.
The inner product is defined by the hermitian metrics
${g_{i\bar{\jmath}}}$ and ${g_{z\bar{z}}}$. That is to say, the
effective action is a bilinear form of the anti-ghost zero modes with
the curvature of the vector bundle {\NU}.
Thus the integral of the anti-ghost zero modes gives a top Chern
class ${c_T}({\NU})$ of the bundle {\NU}. The remaining ghost zero
modes $({{\chi^i}\,,\,{\chi^{z}_{\bar{z}}}})$
with the relation (\ref{eqn:tangen}) govern the tangent space of the
holomorphic instanton space. In other words, they describe the tangent
space of the moduli space of stable maps {\smap}. This moduli space is
defined by a set of the $s$-distinct marked points
$({{x_1},{x_2},\cdots ,{x_s}})$ on ${\Sigma}$ and a holomorphic map
$f$ from the Riemann surface ${\Sigma}$ to the target space M.
The ``stable'' means that a map $f\,;\,\Sigma \rightarrow \mbox{M}$
has no non-trivial first order infinitesimal automorphisms, identical
on M. In other words, it means that the automorphism group of
$({\Sigma\,;\,{x_1}\,\cdots ,{x_s}\,;\,f})$ is finite. (Concretely the
condition means that every component of ${\Sigma}$ of genus 0 (resp.
1) which maps to a point must have at least 3 (resp. 1) marked or
singular points). The moduli space of stable maps to M of curves is
defined as,
\beq
\smap :=
\left\{{
({\Sigma\,;\,{x_1},\cdots ,{x_s}\,;\,f})
}\right\}/\cong  \,\,\,,\label{eqn:teigi}
\eeq
where ``$\cong $'' is the action of the finite automorphism group of
$({\Sigma\,;\,{x_1},\cdots ,{x_s}\,;\,f})$. Also the degree $n$ of the
map $f$ is a non-negative integer determined by the homology class
${H_2}({\mbox{M}\,;\,{\bz}})$ and the homotopy class of the map.

Next we consider the topological gravity part ${S_G}$. For each marked
point ${x_i}$ on ${\Sigma}$, a holomorphic cotangent space
${T^{\ast}_{x_i}}{\Sigma}$ is defined. When one deforms the complex
structure of ${\Sigma}$, the meaning ``holomorphicity'' varies. Then a
complex line bundle ${{\cal L}_{(i)}}$ over the base space {\smap} with
the fiber ${T^{\ast}_{x_i}}{\Sigma}$ is introduced naturally. When one
considers a correlator ${\corr{\cdots {\sigma_{n_i}}({\omega_i})
\cdots }}$
containing a ${n_i}$-th gravitational
descendant ${\sigma_{n_i}}({\omega_i})$ of a cohomology element
${\omega_i}\in {H^{2{q_i}}}({\mbox{M}})$, it contributes after the
path integral as,
\[
{c_1}{{({{\cal L}_{(i)}})}^{n_i}}\,{f^{\ast}}({\omega_i})({x_i})\,\,\,.
\]
Collecting the matter parts and the gravity parts together, we can
write the correlation functions in terms of geometry,
\beqs
{\corr{
{\sigma_{n_1}}({\omega_1}){\sigma_{n_2}}({\omega_2})
\cdots
{\sigma_{n_s}}({\omega_s})
}_{g,n}}=
{\int_{\overline{\cal M}_{g,n}(\mbox{\scriptsize M},n)}}
{c_T}\left({{\NU}_{g,s,n}}\right)\,{\prod^{s}_{i=1}}\,
{{{c_1}({{\cal L}_{(i)}})}^{n_i}}\,{f^{\ast}}({\omega_i})({x_i})\,\,\,,
\eeqs
provided that the following selection rule is satisfied,
\beqs
{\sum^{s}_{i=1}}({{n_i}+{q_i}})=
({\dim \mbox{M}-3})(1-g)+
{\int_{\Sigma}}{f^{\ast}}{c_1}({\mbox{M}})+s\,\,\,.
\eeqs
For our genus one case of the Calabi-Yau $A$(M)-model, the first and
the second terms vanish and then the selection rule depends on the
number of the marked points $s$ only.

Now we express the ${\del_t}{{\sf F}_1}({\bar{t}\rightarrow \infty})$
in geometrical terms as,
\beq
{\del_t}{{\sf F}_{1}}&=&
{\sum^{\infty}_{n=0}}{q^n}
{{\corr{{\sigma_0}({\omega})}}_{1,n}} \hspace{12mm}
({g=1\,\,,\,\,\omega \in {H^{1,1}}({\mbox{M}})}) \nonumber \\
&=&{\sum^{\infty}_{n=0}}{q^n}
\left\{{
{\int_{{\overline{\cal M}_{1,1}}
({\mbox{\scriptsize
M}},n)}}\,{c_T}({\NU_{1,1,n}})\,{f^{\ast}}({\omega})}\right\} \,\,\,.
\label{eqn:chern}
\eeq
Particularly the degree 0 part associated to the {\kae} form $\omega $
can be calculated,
\beqs
{\corr{{\sigma_0}({\omega})}_{1,0}}=
-\frac{1}{24}{\int_{\mbox{\scriptsize M}}}\omega\wedge
{c_{d-1}}({\mbox{M}})\,\,\,.
\eeqs
(This can be used to fix the normalization of the ${{\sf F}_1}$ ).

Finally we make several remarks; the holomorphic maps for ${g\leq 1}$
are isolated only at ${\dim \mbox{M}=3}$. In this case, one can count
the number of the curves embedded in M unambiguously. But the ${\dim
\mbox{M} > 3}$ cases, the anti-ghost zero modes exist and the maps
appear as families. In such cases, the counting the number of the
``individual'' curves no longer lose its meaning and one should
interpret ${\del_t}{{\sf F}_1}({\bar{t}\rightarrow \infty})$ as an
integral (\ref{eqn:chern})
of the top Chern class over the moduli space
${{\overline{\cal M}_{1,1}}({\mbox{M},n})}$.
Also for ${\dim \mbox{M}=3}$ cases, the ${\del_t}{{\sf
F}_{1}}({\bar{t}\rightarrow \infty})$ have contributions from the
rational maps as well as elliptic maps because of the bubbling of the
torus \cite{BCOV}. In our higher dimensional cases $({\dim M > 3})$,
the similar bubbling phenomena are expected to appear. However the
explanation of these possible bubbling needs some new formulation in
moduli spaces of holomorphic instanton families which may have singular or
non-smooth configurations. It remains an open problem.
%%%%%%%%%%%%%%%%%%%%%%%   chotto yasumi
%%%%%~~~~~~~~~~~~~~~~~~~
\section{Conclusion}

\pr
In this article, we have investigated some properties of the
Calabi-Yau d-fold M embedded in ${{CP}^{d+1}}$
 subject to the assumption of the existence of the
mirror symmetries.
We introduce a new quasi-topological field theory
${A^{\ast}}$(M)-model associated to the M. This model is compared to
the $A$(M)-model. By the analysis in the $A{A^{\ast}}$-fusion of these
two models, two point functions of the moduli space associated to the
non-linear sigma model are investigated.
A set of the two point correlators satisfies a set of equations
({$A{A^{\ast}}$-equation}) for the {\kae} manifold M. Because we
switch of all perturbation operators on the topological theory except
for marginal ones ${\ftu{(1)}}$, ${\ftbu{({\bar{1}})}}$ associated
with a {\kae} form of M, the $A{A^{\ast}}$-equation is characterized
by three point functions ${{\Kappa_{l-1}}:=\corr{\ftu{(d-l)}
\Big| \ftu{(1)} \Big| \ftu{(l-1)}}}$ and their conjugates
${{\bar{\Kappa}_{l-1}:=\corr{\ftbu{({\overline{d-l}})} \Big|
\ftbu{({\bar{1}})} \Big| \ftbu{({\overline{l-1}})}}}}$. (The author calculated
these three point functions on the sphere in the previous paper \cite{sugi}).
That is to say, the fusion structure constants in the $A$(M)-model
control the system,
\beqs
&&\ftu{(1)}\cdot \ftu{(l-1)}={\Kappa_{l-1}}\,\ftu{(l)}\,\,\,,\\
&&\ftu{(1)}\cdot \ftu{(d)}=0\,\,\,.
\eeqs
For our Calabi-Yau case, this $A{A^{\ast}}$-equation turns out to be a
non-affine A-type Toda equation. This non-affine property stems
from the nilpotency of the operator products, ${\ftu{(1)}\cdot
\ftu{(d)}}=0$. This nilpotency originates in the vanishing first Chern
class of the Calabi-Yau manifold. For the ${{CP}^{d+1}}$ case, its
first Chern class does not vanish and then the $A{A^{\ast}}$-equation
is an affine Toda equation system. In order to obtain genus one
partition function of the non-linear sigma model described by the
Lagrangian ${L({t\,,\,\bar{t}})}$, we used the data of the two
correlators. By taking an asymmetrical limit ${\bar{t}\rightarrow
\infty}$ and $t$ is fixed, the ${A^{\ast}}$-model part is decoupled
and the above partition function is reduced to that of the
$A$(M)-matter coupled with the the topological gravity at the stringy one
loop level. The coefficients of the ${{\sf F}_1}({\bar{t}\rightarrow
\infty})$ with respect to the indeterminate ${q:={e^{2\pi i t}}}$
represents integrals of the top Chern class of the vector bundle
{\Large ${\nu}$} over the moduli space of the stable curves with
definite degrees $n$.
In this paper we used the mirror conjecture and
our results should be verified by the mathematical methods in enumerative
geometry {\cite{K}}.

%\section*{Acknowledgement}
%\pr
%I would like to express my sincere gratitude to
%Prof.~T.~Eguchi for guidance
%and kindful encouragement throughout my graduate course.
%I also thank Dr.~K.~Hori
%for useful discussions and comments.

%%%%%%%   kokokara %%%%%%%%%%%%%%%%%%
%%%%%%%%%%%%%%%%%%
\section*{Appendix A}
\section*{The $A{A^{\ast}}$-Fusion}

Firstly let us introduce a set of connections
${{\ca}}$,\,${\overline{\ca}}$ defined as,
\beqs
&&{{\ca}_{j\bar{k}}}=\corr{\ftbu{(\bar{k})} \Big| {{\ca}} \Big| \ftu{(j)} }:=
\corr{\ftbu{(\bar{k})} \Big| {{\del}_t} \Big| \ftu{(j)} }\,\,\,,\\
&&{{\bca}_{ j\bar{k}}}=\corr{\ftbu{(\bar{k})} \Big|
{\overline{\ca}}
\Big| \ftu{(j)} }
:=\corr{\ftbu{(\bar{k})} \Big| {{\del}_{\bar{t}}} \Big| \ftu{(j)} }\,\,\,,\\
&&\,\,\,\,\,\,{\del_t}:=\frac{\del}{\del {t}}\,\,\,,\,\,\,
{\del_{\bar{t}}}:=\frac{\del}{\del {\bar{t}}}\,\,\,,
\eeqs
where ${\ftu{(j)}}$, ${\ftbu{({\bar{k}})}}$ are observables of the
$A$-model , ${A^{\ast}}$-model respectively.
Using these connections, we define covariant derivatives,
\beqs
&&{D_t}:={\del_t}-{{\ca}}\,\,\,,\\
&&{\bar{D}_{\bar{t}}}:={\del_{\bar{t}}}
-{\overline{\ca}}\,\,\,.
\eeqs
{}From the path integral representation, the next relation follows,
\beq
\overline{\ca}_{jk} &=& \corr{\ftu{(k)} \Big| \overline{\ca} \Big|
\ftu{(j)}}\nonumber \\
&=& \corr{\ftu{(k)} \Big| {\del_{\bar{t}}} \Big|
\ftu{(j)}}=0
\,\,\,.
\eeq
Also we define a notation,
\[
{\ca}_{jk} :=
\corr{\ftu{(k)} \Big| {\del_t} \Big| \ftu{(j)}} \,\,\,,
\]
and consider a derivative of this
${{\del_{\bar{t}}}{{\ca}_{jk}}}$
with respect to the parameter
${\bar{t}}$.
Then we obtain,
\beq
&&{\del_{\bar{t}}}{{\ca}_{jk}}=
{\del_{\bar{t}}}{{\ca}_{jk}}-{\del_{{t}}}{{\bca}_{jk}}\nonumber \\
&&=
\corr{\ftu{(k)} \left({
{\int_{\Sigma_L}}
\ac{Q_R}{\cm{\widetilde{Q}_L}{\ftbu{(\bar{1})}}}
}\right)
\Big|
\left({
{\int_{\Sigma_R}}
\ac{\widetilde{Q}_R}{\cm{Q_L}{\ftu{(1)}}}
}\right) {\ftu{(j)}} }\nonumber \\
&&\hspace{5mm}-
\corr{{\ftu{(k)}} \left({
{\int_{\Sigma_L}}
\ac{Q_R}{\cm{\widetilde{Q}_L}{\ftu{(1)}}}
}\right) \Big|
\left({
{\int_{\Sigma_R}}
\ac{\widetilde{Q}_R}{\cm{Q_L}{\ftbu{(\bar{1})}}}
}\right) \ftu{(j)} }\,\,\,\nonumber \\
&& = \corr{\ftu{(k)} \left({
{\int_{\Sigma_L}}
\ftbu{(\bar{1})} }\right) \Big|
\left({
{\int_{\Sigma_R}}{\del_z}{\bar{\del}_{\bar{z}}} \ftu{(1)}
}\right) \ftu{(j)} }\nonumber \\
&&\hspace{5mm}
- \corr{\ftu{(k)} \left({
{\int_{\Sigma_L}}
\ftu{({1})} }\right) \Big|
\left({
{\int_{\Sigma_R}}{\del_z}{\bar{\del}_{\bar{z}}} \ftbu{(\bar{1})}
}\right) \ftu{(j)} }\,\,\,,\label{eqn:deri1}
\eeq
by using the relations,
\beqs
&&\ac{Q_L}{\widetilde{Q}_L}={\del_z}\,\,\,,\\
&&\ac{{Q}_R}{\widetilde{Q}_R}={\bar{\del}_{\bar{z}}}\,\,\,.
\eeqs
The first term in ({\ref{eqn:deri1}}) is rewritten as,
\beqs
&& \corr{{\ftu{(k)}} \left({
{\int_{\Sigma_L}}
\ftbu{(\bar{1})} }\right) \Big|
\left({
{\int_{\Sigma_R}}
{\del_z}{\bar{\del}_{\bar{z}}}{\ftu{(1)}}
}\right){\ftu{(j)}}} \\
&&=-\corr{\ftu{(k)} \left({
{\int_{\Sigma_L}}
{\ftbu{(\bar{1})}} }\right) \Big|
\left({
{\oint_{C}}{\del_n}
{\ftu{(1)}}
}\right) \ftu{(j)}}\,\,\,\\
&& =-\corr{\ftu{(k)} \left({
{\int_{\Sigma_L}}
{\ftbu{(\bar{1})}} }\right) \Big|
\left({H\,
{\oint_{C}}
{\ftu{(1)}} }\right) \Big| \ftu{(j)}
} \,\,\,,
\eeqs
where $C$ is the boundary of ${\Sigma_R }$ and ${\del_n}$ is the
normal derivative along the cylindrical direction of ${\Sigma}$ and
$H$ is a hamiltonian along this direction of ${\Sigma}$,
\[
{{\del_n}\ftu{(i)}}=\cm{H}{\ftu{(i)}}\,\,\,.
\]
Furthermore the above formula is re-expressed as,
\beqs
&& -\corr{\ftu{(k)} \left({
{\int_{\Sigma_L}}
{\ftbu{(\bar{1})}} }\right) \Big|
\left({H\,
{\oint_{C}}{\ftu{(1)}} }\right) \Big| \ftu{(j)}
} \\
&&=-{\int^{T}_{0}}d{\tau} \corr{\ftu{(k)} \Big| \left({
{\oint_{C'}}{\ftbu{(\bar{1})}} }\right)
H\,{e^{-{\tau}H}}\left({
{\oint_{C}}{\ftu{(1)}}}\right) \Big| \ftu{(j)}
} \,\,\,,
\eeqs
where the integration ${\int}d{\tau}$ is over the length of the left
cylinder ${\Sigma_L}$.
Finally taking a long-cylindrical limit ${T\rightarrow \infty }$,
the ${{\del_{\bar{t}}}{{\ca}_{jk}}}$ becomes,
\beqs
&&  \corr{\ftu{(k)} \Big| \left({
{\oint_{C'}}
{\ftbu{(\bar{1})}} }\right) {e^{-TH}}
\left({{\oint_{C}}
{\ftu{(1)}}
}\right)\Big| \ftu{(j)}
}_{({T\rightarrow \infty})} \\
&& - \corr{\ftu{(k)} \Big| \left({
{\oint_{C'}}
{\ftu{(1)}}
 }\right) {e^{-TH}}
\left({{\oint_{C}}
{\ftbu{(\bar{1})}}
}\right)\Big| \ftu{(j)}
}_{({T\rightarrow \infty})} \\
&&={\beta^2}{{\left({\bar{C}_{\bar{t}} }\right)}^{n}_{j}}
{{\left({{C}_{{t}} }\right)}_{nk}}
-{\beta^2}
{{\left({{C}_{{t}}}\right)}^{n}_{j}
\left({
{\bar{C}_{\bar{t}}} }\right)}_{nk}   \,\,\,,
\eeqs
where ${C_t}$, ${\bar{C}_{\bar{t}}}$ are three point couplings of the
$A$-model, the ${A^{\ast}}$-model respectively and are defined as,
\beqs
&&{{\left({{C_t}
}\right)}_{lm}}:=
\corr{\ftu{(m)} \Big| \ftu{(1)} \Big| \ftu{(l)}}\,\,\,,\\
&&{{\left({{\bar{C}_{\bar{t}}}
}\right)}_{\bar{l}\bar{m}}}:=
\corr{\ftbu{(\bar{m})} \Big| \ftbu{(\bar{1})} \Big| \ftbu{(\bar{l})}}\,\,\,.
\eeqs
Also the symbol ${{\left({\bar{C}_{\bar{t}}
}\right)}^{k}_{j}}$ means that,
\[
{{\left({\bar{C}_{\bar{t}}
}\right)}^{k}_{j}} =
{{\gs}_{j\bar{n}}}
{{\left({\bar{C}_{\bar{t}}
}\right)}^{\bar{n}}_{\bar{m}}}{{\gs}^{\bar{m} k}} \,\,\,.
\]
For the hermitian metric ${{\gs}_{l\bar{m}}}$ and the covariant
derivative ${{D_t}={\del_t}-{\ca}}$, the next relation is understood,
\[
{D_t}{{\gs}_{l\bar{m}}}=0\,\,\,.
\]
That is to say, the connection can be written as,
\[
{{\ca}^{l}_{j}}=-{{\gs}_{j\bar{n}}}
{({ {\del_t}{{\gs}^{-1}} })}^{\bar{n} l} \,\,\,.
\]
Now we raise a holomorphic index of the ${{\ca}_{jk}}$ by multiplying
the topological metric ${{\eta}^{lk}}$ and obtain a final result,
\beqs
{\del_{\bar{t}}}{{\ca}^{l}_{j}}&=&
-{\del_{\bar{t}}} {\left({
{\gs}\,{\del_t}{{\gs}^{-1}}
}\right)}^{l}_{j}  \\
&=& -{\beta^2} {\cm{C_t}{{\gs}\,{\bar{C}_{\bar{t}}}\,{{\gs}^{-1}} }}^{l}_{j}
\,\,\,.
\eeqs

\section*{Appendix B}
\section*{The Three Point Functions on the Sphere}

In this appendix B, we derive the three point functions on the sphere
of the d-fold ({\ref{eqn:CY}}) under the mirror symmetry. (More detail
can be seen in \cite{sugi}).

%%%%%%%%%%%%%%%%%%%%%%%%bbbbbbbbbbbbbbbbbb appendixbbegin

Firstly we take notice of the Hodge structure of the primary
horizontal subspace ${H^d}$(W) of the mirror manifold W paired with M.
The deformation of the complex structure of $W$ is controlled by the
variation of the Hodge decomposition of ${H^d}(W)$ and
its information is given by the period matrix $P$ of $W$.
This period matrix is defined by using homology d-cycles
${\gamma_j}\in {H_d}(W)$, cohomology elements,
\beqs
&&{\alpha_i}:={\Theta^{i}_{z}}{\Omega}\in
{{\cal F}^{d-i}}={H^{d,0}}\oplus {H^{d-1,1}}\oplus \cdots \oplus
{H^{d-i,i}}\,\,\,,
\\
&&N:=d+2\,\,,\,\,z:={({N\psi})^{-N}}\,\,,\,\,{\Theta_{z}}:=z\cdot
\frac{d}{d\,z}\,\,\,.
\eeqs
Its matrix elements ${P_{ij}}\,\,(0\leq i \leq d\,,\,0\leq j \leq d)$
are expressed as,
\[
{P_{ij}}:={\int_{\gamma_j}}{\alpha_i}\,\,\,.
\]
The ${\alpha_0}={\Omega}$ is a globally defined
nowhere-vanishing holomorphic
d-form of $W$ and can be expressed for the Fermat type
hypersurface ${\ps}$ as, % {\cite{LSW,G,BG}},
\beqs
&&\Om :={\int_{\gamma}}{\frac{d\mu}{\ps}}\,\,\,,\\
&&d\mu :={\sum^{d+2}_{a=1}}{{(-1)}^{a-1}}
{X_a} d{X_1}\wedge d{X_2}\wedge \cdots \wedge
\widehat{d{X_a}} \wedge \cdots \wedge d{X_{d+2}}\,\,\,,
\eeqs
where ${\gamma}$ is a small one-dimensional cycle
winding around the hypersurface defined as a zero locus of ${\ps}$.
Using this explicit formula of the ${\Omega}$, we obtain a
differential equation satisfied by ${P}$,
\beqs
&&{\th{z}}\,P (z)={\widetilde{C}_z}\,P (z)\,\,\,,\\
&&{\widetilde{C}_z}:=\left(
\begin{array}{cccccc}
0 & 1 &  &  &  &  \\
  & 0 & 1 &  &  &   \\
 &  & \ddots  & \ddots  &  &  \\
  &   &  & 0  & 1  &  \\
  &   &  &    & 0  & 1 \\
{\sigma_{N-1}}  & {\sigma_{N-2}}  & {\sigma_{N-3}} & \cdots   &
{\sigma_{2}} & {\sigma_{1}}
\end{array}
\right) \,\,\,,\\
&&{\sigma_{m}}:={\displaystyle \frac{{N^N} z}{1-{N^N} z}}
\begin{array}[t]{c}
{\displaystyle {\mathop{{\sum}}}}\\
{\scriptstyle 1\leq {n_1} < {n_2} < \cdots < {n_m} \leq N-1}
\end{array}
\frac{n_1}{N}\cdot
\frac{n_2}{N}\cdot
\cdots
\frac{n_m}{N} \,\,\,.
\eeqs
Each component ${P_{ij}}$ of ${P}$ is obtained,
\beqs
&&{P_{0j}}={{\vpi}_j} ({z}):=
{\sum^{j}_{l=0}}\,\frac{1}{l!}\,
{\left({\frac{\log z}{2\pi i}}\right)}^{l}
\times {\sum^{\infty}_{m=0}}{b_{j-l,m}}\cdot {z^m}\,\,\,,\\
&&{b_{n,m}}:=\frac{1}{n!} {\left({\frac{1}{2\pi i}\cdot
\frac{\del}{\del\rho}}\right)}^{n}
\left\{{
\frac{\Gamma ({N({m+\rho})+1})}{\Gamma ({N{\rho}+1})}\cdot
%{\prod^{d+2}_{i=1}}
{{\left[
\frac{\Gamma ({{\rho}+1})}{\Gamma ({m+\rho +1})}
\right]}^N}}\right\} {\Biggr|_{\rho =0}} \,\,\,,\\
&&{P_{ij}}={\int_{\gamma_j}}{\Theta^{i}_{z}}\Om ={\Theta^{i}_{z}}{\varpi_j}
\,\,\,.
\eeqs
In the equation, the derivative ${\Theta_{z}}$ induces the variation
of the complex structure of W and the structure of the Hodge
decomposition is deformed.
Also each entry of the $l$-th row of the $P$ belongs to the ${{\cal
F}^{d-l}}$.
So the equation describes a linear representation of the infinitesimal
deformation of the complex structure of W.

Next by introducing a new variable ${t={\omega_1}(z)}$, we rewrite the
above differential equation on the upper triangular matrix ${\Phi}$,
\beqs
&&{\del_{t}}\Phi (t)={C_t}\,\Phi  (t)\,\,\,,\,\,\,
{\Phi_{lm}}:={\int_{\gamma_m}}{\tilde{\alpha}_l}(t)\,\,\,,\\
&&{C_t}:=\left(
\begin{array}{ccccccc}
 0 & {\Kappa_0} &  &  &  &  & \mbox{\Large $O$} \\
   & 0 & {\Kappa_1}  &  &  &  &  \\
   &   & 0 & {\Kappa_2}  &  &  &  \\
   &   &   & \ddots  & \ddots &  &  \\
   &   &   &   & 0  & {\Kappa_{d-2}} &  \\
   &   &   &   &   & 0 & {\Kappa_{d-1}} \\
\mbox{\Large $O$} &   &   &   &   &   & 0
\end{array}
\right) \,\,\,,\\
&&{\tilde{\alpha}_0}(t):=\frac{1}{\varpi_0}{\alpha_0}\,\,\,,\\
&&{\tilde{\alpha}_l}(t):=
\frac{1}{{\Kappa}_{l-1}}{\del_{t}}
\frac{1}{{\Kappa}_{l-2}}{\del_{t}}\cdots {\del_{t}}
\frac{1}{{\Kappa}_{1}}{\del_{t}}
\frac{1}{{\Kappa}_{0}}{\del_{t}}
\left({\frac{\alpha_0}{\varpi_0}}\right)
\,\,\,,\,\,\,({1\leq l \leq d})\,\,\,,\\
&&{{\Kappa}_0}:={\del_{t}}
\,{\omega_1}=1\,\,\,,\,\,({t\equiv {\omega_1}})\\
&&{{\Kappa}_m}:=
{\del_{t}}\frac{1}{{\Kappa}_{m-1}}{\del_{t}}
\frac{1}{{\Kappa}_{m-2}}{\del_{t}}\cdots {\del_{t}}
\frac{1}{{\Kappa}_{1}}{\del_{t}}
\frac{1}{{\Kappa}_{0}}{\del_{t}}\,{\omega_{m+1}}
\,\,\,,\,\,\,({1\leq m \leq d-1})\,\,\,.
\eeqs
{}From this equation, we can read the action of the differential
operator ${\del_{t}}$
on the cohomology basis ${{\tilde{\alpha}_{l}}(t)}$,
\beqs
&&{\del_{t}}{\tilde{\alpha}_{j-1}}(t)={\Kappa_{j-1}}(t)\,
{\tilde{\alpha}_{j}}(t)\,\,\,,\,\,\,({1\leq j \leq d})\,\,\,,\\
&&{\del_{t}}{\tilde{\alpha}_{d}}(t)=0\,\,\,,\\
&&{\Kappa_{j-1}}(t)={\del_{t}}{\Phi_{j-1\,,\,j}}\,\,\,,\,\,\,({1\leq j \leq d})
\,\,\,.
\eeqs
Because the ${t(z)}$ is a mirror map and couples with a {\kae} form
${J:=t\cdot e}$ of M, the derivative ${\del_t}$ can be interpreted as
an insertion of the operator ${\ftu{(1)}}$ associated to the {\kae}
form $e$ in the $A$(M)-model terms.
%%%%%%%  koko
We translate these relations into
the operator structures of the A(M)-model,
\beqs
&&{{\cal O}^{(1)}}{{\cal O}^{({j-1})}}=
{\Kappa_{j-1}}(t){{\cal O}^{(j)}}\,\,\,,\,\,\,({1\leq j \leq d})\,\,\,,\\
&&{{\cal O}^{(1)}}{{\cal O}^{({d})}}=0\,\,\,,
%&&\corr{{{\cal O}^{(i)}}{{\cal O}^{(j)}}}=
%{\delta_{i+j,d}}{\Eta_{ij}}\,\,\,,
\eeqs
for A(M)-model operators ${{{\cal O}^{(i)}}\in {H^{i,i}_{J}}(M)}\,,\,
({1\leq i \leq d})$.
%\newpage
The above operator product structure of the A(M)-model observables
is meaningful when one defines correlation functions in the following
way,
\beqs
&&\corr{{{\cal O}^{(1)}}{{\cal O}^{(j-1)}}\cdots }
:= \int {\cal D}[{X,\chi , \rho}]\,
{{\cal O}^{(1)}}{{\cal O}^{(j-1)}}\cdots {e^{-{L_A}}}\,\,\,,\\
&&\,\,\,\,\,\,{L_A}:=-\sqrt{-1}
{\int_{\Sigma}}{X^{\ast}}(e)+  {\int_{\Sigma}}{d^2}z \,
\ac{Q^{(+)}}{V^{(-)}}\,\,\,,
\eeqs
where ${Q^{(+)}}$ is a BRST charge of the A(M)-model and ${V^{(-)}}$
is given in ({\ref{eqn:modela}}).
Thus we can obtain the fusion couplings ${\{{\Kappa_l}\}}:
=\corr{{\ftu{(1)}}{\ftu{(l)}}{\ftu{(d-l-1)}}}$,
\beqs
&&{\Kappa_0}=1\,\,\,,\\
&&{\Kappa_{l}}=
{\del_{t}}\frac{1}{\Kappa_{l-1}}
{\del_{t}}\frac{1}{\Kappa_{l-2}}{\del_{t}}\cdots
{\del_{t}}\frac{1}{\Kappa_{2}}
{\del_{t}}\frac{1}{\Kappa_{1}}
{\del_{t}}\frac{1}{\Kappa_{0}}
{\del_{t}}\,{S_{l+1}}({{\tilde{x}_1}\,,\,{\tilde{x}_2}\,,\,\cdots \,,\,
{\tilde{x}_{l+1}}})\,\,\,,\,\,\,({1\leq l \leq d-1})\,\,\,,\\
&&{\tilde{x}_{n}}:=\frac{1}{n!}{{D}^{n}_{\rho}}\log {\hat{\varpi}_0}
({z\,;\,\rho})
{\Big|_{\rho =0}}\,\,\,,\,\,\,
{{\dc}_{\rho}}:=\frac{1}{2\pi \,i }\cdot \frac{\del}{\del \rho}\,\,\,,\\
&&{\hat{\vpi}_0} ({z,\rho}):= \sum^{\infty}_{m=0}
\frac{\Gamma ({N({m+\rho})+1})}{\Gamma ({N{\rho}+1})}\cdot
{{\left[{
\frac{\Gamma ({{\rho}+1})}{\Gamma ({{m+\rho}+1})}
}\right]}^{N}}
\cdot {z^{m+\rho}}
\,\,\,,
\eeqs
where the function ``${S_n}$'' is
the Schur function defined as the coefficients
in the following expansion,
\[
{\sum^{\infty}_{n=0}}{S_n}({x_1},{x_2},\cdots ,{x_n}) {u^n}:=
\exp \left({{\sum^{\infty}_{m=1}}{x_m}{u^m}}\right) \,\,\,.
\]
We write down expressions of these couplings ${\Kappa_{l}}$ in a series
with respect to a parameter ${{q:={e^{2\pi i\,t}}}\,\,,\,\,
t={S_1}({\tilde{x}_1})={\tilde{x}_1}}$,
\beqs
&&{\Kappa_l}=1+{\alpha_l}\,q+\mbox{\large $O$}({q^2})\,\,\,,\\
&&{\alpha}_l = {N!}\times
\,
\left({
{\displaystyle
{\tilde{A}}_{d+1-l} -
\sum^{N}_{l=2} \frac{N}{l}
}
}\right)
\,\,\,\,,\\
&&{{\tilde{A}}_m} :=
\begin{array}[t]{c}
{\displaystyle {\mathop {{\sum}}}}\\
{\scriptstyle 1 \leq {m_1}<{m_2}< \cdots  <{m_n} \leq N-1 }
\end{array}
\frac{N-{m_1}}{m_1}\cdot \frac{N-{m_2}}{m_2}\cdot \cdots
\frac{N-{m_n}}{m_n} \,\,\,.
\eeqs

%%%%%%%%%%%%%%%%%%aaaaaaaaaaaaaaaaaaaaaaaaaaaaaaaaaaaaappendixbens
\section*{Appendix C}
\section*{The Explicit Forms of the {\kae} Potential}

We calculate several explicit forms of the {\kae} potential
${\cal K}$ in some lower dimensions for our case.

\[
\begin{array}{l}
\mbox{
\begin{tabular}{|c|l|} \hline
\mbox{dimension}  &
 ${e^{-{\cal K}}}\cdot {({{\vpi_0}{\bar{\vpi}_0}})}^{-1}$ \\ \hline
1 & $t-\bar{t}$ \\ \hline
2 & ${\displaystyle \frac{1}{2}{(t-\bar{t})}^2 +
\frac{1}{2}\left({ {{\cal D}_{\rho}}t+
\overline{{{\cal D}_{\rho}}t}
}\right){{}_{\rho =0}}
}$ \\ \hline
3 & ${\displaystyle \frac{1}{6}{(t-\bar{t})}^3 +
\frac{1}{2}({t-\bar{t}}) \left({ {{\cal D}_{\rho}}t+
\overline{{{\cal D}_{\rho}}t}
}\right){{}_{\rho =0}}
+\frac{1}{6} \left({ {{\cal D}^2_{\rho}}t-
\overline{{{\cal D}^2_{\rho}}t}
}\right){{}_{\rho =0}}
}$ \\ \hline
4 & ${\begin{array}{l}
{\displaystyle \frac{1}{24}{(t-\bar{t})}^4 +
\frac{1}{4}{({t-\bar{t}})}^2  \left({ {{\cal D}_{\rho}}t+
\overline{{{\cal D}_{\rho}}t}
}\right){{}_{\rho =0}}
+\frac{1}{6} ({t-\bar{t}})
 \left({ {{\cal D}^2_{\rho}}t-
\overline{{{\cal D}^2_{\rho}}t}
}\right){{}_{\rho =0}}
} \\
{\displaystyle  +\frac{1}{24}
 \left({ {{\cal D}^3_{\rho}}t+
\overline{{{\cal D}^3_{\rho}}t}
}\right){{}_{\rho =0}}
 +\frac{1}{8}
 {{\left({ {{\cal D}_{\rho}}t+
\overline{{{\cal D}_{\rho}}t}
}\right)}^{2}}{{}_{\rho =0}}
}
\end{array}}$
\\  \hline
\end{tabular}
}\\
\mbox{}
\end{array}
\]
%\eeqs
The ${\rho}$-derivatives are defined as,
\beqs
{{\cal D}^{m}_{\rho}}t :=
{{\left({\frac{1}{2\pi i}\cdot \frac{\del}{\del \rho}}\right)}^{m+1}}
\cdot {\!}\log {\hat{\varpi}_0}({z\,;\,\rho})\,\,\,,
\eeqs
and the ``bar'' means the complex conjugate of them.
The above examples coincide with the well-known results of the torus
({\,${d=1}$\,}), the K3 surface ({\,${d=2}$\,}) and the Quintics
({\,${d=3}$\,}). In the B(W)-model case, the {\kae} potential can be defined
by using a holomorphic $d$-form $\Om $ and an anti-holomorphic $d$-form
$\bar{\Om}$ as,
\[
{e^{-{\cal K}}}={\int_W}{\Om}\wedge {\bar{\Om}}\,\,\,.
\]
On the other hand, this $d$-form can be expanded by a set of dual basis of
the homology cycles ${\{{\gamma_l}\}}$,
\beqs
&&{\Om}={\sum^{d}_{l=0}}\left({{\int_{\gamma_l}}\Om }\right)
{\gamma^{\ast}_{l}}={\sum^{d}_{l=0}}({\vpi_{l}}) \cdot {\gamma^{\ast}_{l}}
\,\,\,,\\
&&{\int_{\gamma_m}}{\gamma^{\ast}_{l}}={\delta_{l,m}}\,\,\,\,,\,\,\,\,
{\gamma_l}\in {H_d}({W\,;\,{\bz}}) \,\,\,.
\eeqs
We suppose that intersection numbers of these cycles are given by,
\beqs
&&{{\bf I}_{l,m}}:={\int_W}
{\gamma^{\ast}_{l}}\wedge {\gamma^{\ast}_{m}}\,\,\,,\\
&&{\bf I}= \left(
\begin{array}{ccccc}
\mbox{\Large $O$} & & & & {(-1)}^{d} \\
 & & & {(-1)}^{d-1} & \\
 & & \cdots  & & \\
 & {(-1)}^{1} & & &  \\
{(-1)}^{0} & & & & \mbox{\Large $O$}
\end{array}
\right)
\,\,\,.
\eeqs
When one performs the monodromy transformation around the point ${z=0}$
as \\
${z\rightarrow \exp ({2\pi \sqrt{-1}\,  n}) \cdot z }$, the functions
${\vpi_{l}}$ transform,
\beqs
&&\vpi :={{}^{t}}
\left(
\begin{array}{cccc}
\vpi_{0} & \vpi_{1} & \cdots & \vpi_{d}
\end{array}
\right) \,\,\,,\\
&&\vpi \rightarrow {T_n}\cdot \vpi \,\,\,\,\,\,\,\,\,\,\,
({z\rightarrow \exp ({2\pi \sqrt{-1} n}) \cdot z })\,\,\,,\\
&&{T_n}:={T^n_1}\,\,\,,\\
&&{T_1}:=\exp \left(
\begin{array}{cccccc}
0 &  &  &  &  & \mbox{\Large $O$}\\
1 & 0  &  &  &  & \\
  & 1  & 0  &  &  &  \\
  &    & \ddots & \ddots  &  &   \\
  &    &        &   1     &  0   &  \\
\mbox{\Large $O$}  &    &        &         &  1   & 0
\end{array}
\right)
 \,\,\,.
\eeqs
Then the intersection matrix ${\bf I}$ is affected by this matrix,
\[
{\bf I}\rightarrow {{}^{t}}{T_n}{\bf I}{T_n}={\bf I}\,\,\,.
\]
So we presume that we have an appropriate set of homology
cycles.

Using this assumption, we obtain the {\kae} potential in the monodromy
invariant form,
\beqs
&&{g_{0\bar{0}}}={e^{-{\cal K}}}\\
&&=\left({{\varpi_{0}}{\bar{\varpi}_{0}} }\right)\,
{\sum^{d}_{a=0}} {\om{a}}\omb{d-a}\cdot {(-1)}^{d-a}  \\
&&=\left({{\varpi_{0}}{\bar{\varpi}_{0}} }\right)\,
{S_d}\left({ \, {\tilde{z}_1}\,,\,
 {\tilde{z}_2}\,,\,
\cdots \,,\,
 {\tilde{z}_d}\,
 }\right) \,\,\,,\\
&&{\tilde{z}_m}:={\tilde{x}_m}+{{(-1)}^{m}}{\overline{\tilde{x}}_m}\,\,\,,\\
&&{\tilde{x}_n}:=\frac{1}{n!} {{\cal D}^{n}_{\rho}}\log
{\hat{\varpi}_0}({z\,;\,\rho}) {\big|}_{\rho=0} \,\,\,.
\eeqs

\section*{Appendix D}
\section*{Derivation of the Parameters ${\alpha}$ and ${\beta}$}

\pr
In deriving the unknown parameters ${\alpha}$ and ${\beta}$, we need
the information about the behaviour of the period ${\varpi_0}$ and
the mirror map $t$ around the points ${\psi \sim 0}$ or ${\psi \sim
\infty}$.
Firstly the period ${\varpi_0}$ can be expressed as,
\beqs
{\varpi_0}({\psi})&=&{\sum^{\infty}_{n=0}}
{\displaystyle
\frac{\Gamma {\!}({Nn+1})}{{\left[{\Gamma {\!}(n+1)}\right]}^N}\cdot
\frac{1}{{({N\psi})}^{Nn}}
}\,\,\,,\,\,\,({|{\psi}|}>1)\,\,\,\\
&=&\frac{-1}{N}{\sum^{N-1}_{r=1}}
{{\left({
{\displaystyle  \frac{{\tilde{\alpha}^r}-1}{2\pi i}}
}\right)}^{N-1}}\cdot {{({N\psi})}^{r}}\\
&&\times {\sum^{\infty}_{m=0}}
{\displaystyle
\frac{{\left[{\Gamma {\!}
\left({m+\frac{r}{N}}\right)}\right]}^N}
{\Gamma {\!}\left({Nm+r}\right)}\cdot
{{({N\psi})}^{Nm}}
}\,\,\,\,,\,\,\,({{|{\psi}|}<1 })\,\,\,,
\eeqs
where ${\tilde{\alpha} :=\exp \left({\frac{2\pi i}{N}}\right)}$.
It behaves as,
\beq
\left\{
\begin{array}{cclcl}
{\varpi_0} & \sim & {\psi^{1}} & & ({\psi \sim 0})\,\,\,,\\
{\varpi_0} & \sim & \mbox{regular} & & ({\psi \sim \infty})\,\,\,.
\end{array}
\right.
\eeq
Secondly we can write the mirror map ${t({\psi})}$ as,
\beqs
2\pi i \cdot t({\psi})&=&
N\log \frac{1}{{({N\psi})}}\\
&&+N {\sum^{\infty}_{n=1}}
{\displaystyle
\frac{\Gamma {\!}\left({Nn+1}\right)}{{\left[{\Gamma {\!}
\left({n+1}\right)}\right]}^N}\cdot
\left\{{\Psi \left({Nn+1}\right)- \Psi (n+1)
}\right\}\cdot \frac{1}{{({N\psi})}^{Nn}}
}\,\,\,,\,\,\,({{|{\psi}|}>1})\\
&=& -
\frac{\displaystyle {\sum^{N-1}_{r=1}}{\tilde{\alpha}^r}
{ {({{\tilde{\alpha}^r}-1})}^{N-2}}
\cdot {{({N\psi})}^{r}}{\xi_r}({\psi})}
{\displaystyle {\sum^{N-1}_{s=1}}{{({{\tilde{\alpha}^s}-1})}^{N-1}}
\cdot {{({N\psi})}^{s}}{\xi_s}({\psi})
}
\,\,\,,\,\,\,({{|{\psi}|}<1})\,\,\,,\\
&&{\xi_r}({\psi}):={\sum^{\infty}_{m=0}}
{\displaystyle
\frac{{\left[{
\Gamma {\!}\left({m+\frac{r}{N}}\right)
}\right]}^N}{\Gamma {\!}\left({Nm+r}\right)}\cdot {{({N\psi})}^{Nm}}
}\,\,\,.
\eeqs
{}From this formula, we can write asymptotic behaviours of $t$,
\beqs
\left\{
\begin{array}{cclcl}
t & \sim & \mbox{regular} & & ({\psi \sim 0})\,\,\,,\\
t & \sim & \log {\psi^{-N}} & & ({\psi \sim \infty})\,\,\,.
\end{array}
\right.
\eeqs
In the ${\psi \sim 0}$ case, the ${{\sf F}_1}$ tends to a form,
\[
{{\sf F}_1} \sim \frac{1}{2}\log \left\{{\psi^{\alpha -v}}\right\}\,\,\,.
\]
By postulating the regularity, we can obtain the ${\alpha}$,
\beq
\alpha =v\,\,\,.
\eeq
Also the ${{\sf F}_1}$ behaves as,
\[
{{\sf F}_1} \sim \frac{1}{2} \log \left\{{\psi^{\alpha +N\beta +u}
}\right\}\,\,\,,
\]
in the ${\psi \rightarrow \infty}$ limit.
On the other hand, by the consideration of the large radius limit,
this ${{\sf F}_1}$ turns to be a formula,
\[
{{\sf F}_1}\sim \frac{1}{2} \log \left\{{
\psi^{N\cdot \frac{1}{12}\cdot {N_{d-1}}}
}\right\}\,\,\,.
\]
So the ${\beta}$ can be obtained,
\beq
\beta =\frac{1}{12}\cdot {N_{d-1}}-\frac{u+v}{N} \,\,\,.
\eeq

%%%%%%%%%%%%%bbbbbbbbbbbbbbbbbbbbbb

\end{document}